\documentclass[a4paper, 10pt, journal, twocolumn]{IEEEtran}

\usepackage[utf8]{inputenc}
\usepackage[english]{babel}
\usepackage{amsfonts}
\usepackage[cmex10]{amsmath}
\usepackage{cases}
\usepackage{cite}
\usepackage{graphicx}
\usepackage{subcaption}
\usepackage{epstopdf}
\usepackage{algorithm}
\usepackage{algorithmic}
\usepackage[svgnames]{xcolor}

\begin{document}
% Title page
\title{A concave-convex approach to bit depth allocation for linear MMSE estimation in wireless sensor networks with correlated noise}
\author{Fernando de la Hucha Arce, 
Marian Verhelst %~\IEEEmembership{Senior Member,~IEEE,} 
and 
Alexander Bertrand%~\IEEEmembership{Senior Member,~IEEE}
\thanks{This research work was carried out at the ESAT Laboratory of KU Leuven in the frame of Research Fund KU 
Leuven C14/16/057, FWO nr. G.0931.14, the European Research Council (ERC) under the European Union's Horizon
2020 research and innovation programme (grant agreements No. 802895 and No. 715037).
The scientific responsibility is assumed by the authors. 

The authors are with the Dept. of Electrical Engineering (ESAT), Kasteelpark Arenberg 10, 3001 Leuven, Belgium,
email: \{fernando.delahuchaarce,
marian.verhelst,
alexander.bertrand\}@kuleuven.be.}}
%Copyright \copyright \(\,\) 2020 IEEE. 
%Personal use is permitted, but republication/redistribution requires IEEE permission.}}
\maketitle

% Abstract
\begin{abstract}
Nodes in a wireless sensor network (WSN) are generally powered by 
batteries, and therefore energy efficiency is crucial.
The bit depth used to quantize the sensor signal samples heavily influences energy consumption, as it strongly impacts 
the amount of information to be transmitted between the sensor nodes.
Bit depth allocation problems seek to assign a bit depth to each sensor signal to optimize network resources, such
as energy or bandwidth, together with algorithmic performance.
We study a general bit depth allocation problem for linear minimum mean squared error (MMSE) estimation in WSNs with spatially correlated noise and a communication energy model depending linearly on the sensor bit depths.
This model follows from common configurations in digital communications with orthogonal multiple-access schemes.
After relaxation to real variables, we formulate the resulting non-convex problem as a difference of convex (DC) problem, and develop an algorithm based on the concave-convex procedure (CCP) to find a local optimum.
We show that the convex problem involved in each CCP iteration can be efficiently solved with a worst-case complexity that is linear in the number of sensors.
Additionally, we show that our approach is more computationally efficient than previously proposed bit depth allocation algorithms and generic optimization methods, and it achieves significant energy savings.
\end{abstract}

% Keywords

\section{Introduction} \label{sec:Introduction}
% Wireless sensor networks
A wireless sensor network (WSN) consists of a set of nodes which are able to use their sensors to gather 
information from the environment and to exchange this data over wireless communication links \cite{Trigka25}.
The goal of the network is generally to extract information about a physical phenomenon from the sensor data
collected by the nodes.

% Energy efficiency related to bit depth for quantization
The energy budget of the sensor nodes is often limited since they are typically powered by batteries, which may be difficult to replace or recharge.
As a consequence, energy efficiency is crucial in the design of algorithms for WSNs \cite{AnastasiEnergy09, Rault2014}.
In particular, data exchange through wireless communications is generally more expensive than data processing at
the sensor nodes \cite{AnastasiEnergy09}.
One parameter that has a significant effect on the amount of data to be exchanged between the sensor nodes is 
the bit depth used to quantize the sensor signal samples.
The goal of bit depth allocation is to choose the bit depth of each sensor signal to optimize one (or several) network 
resources, such as energy and bandwidth, together with algorithmic performance.
As we discuss next, a common formulation in the literature is the minimization of the total communication energy
consumption while respecting a performance constraint depending on the task assigned to the WSN.
An alternative formulation is the maximization of performance of the WSN's task while respecting a bandwidth
constraint.
In this paper, we focus on the first formulation and on the task of multi-channel signal estimation, which has applications in, e.g., speech enhancement for wireless acoustic sensor networks \cite{Doclo07, Bertrand09-65, Cornelis11, Markovich13, Hassani15-68, Didier26} and artifact removal in neuro-sensor networks \cite{Bertrand14-125, Narayanan20, Tang20}.

% Related works.
\subsection{Related work} \label{subsec:RelatedWork}
Bit depth allocation has been studied for several estimation algorithms, network topologies and communication 
energy models, often under some assumptions that simplify the resulting optimization problem.
We begin with the most common topology for WSNs in the literature of bit depth allocation, namely the star topology,
where the sensor nodes send their signal samples to a fusion centre (FC), which is in charge of performing the
estimation task.
% Star topology and simplified signal models.
In this context, the work presented in \cite{GoldsmithDigComm06} finds an analytical solution to a bit depth allocation
problem for a WSN whose task is parameter estimation through the best linear unbiased estimator (BLUE).
However, the authors assume that the noise sensor signal components are spatially uncorrelated, which precludes its 
use in many practical applications where noise is spatially correlated, e.g., when localized noise sources are present.
The same assumption about spatially uncorrelated noise is taken in \cite{Sani18}, where the authors study the 
minimization of the mean squared error (MSE) subject to a bandwidth constraint when using the linear minimum mean
squared error (MMSE) estimator.
The effect of noise correlation is studied in \cite{Krasnopeev05}, although for a suboptimal BLUE estimator, as it
is not optimized to take the presence of quantization noise into account.
Besides, all sensors are assumed to be active, i.e. their allocated bit depth is assumed to never be zero.
In many cases, especially when the number of sensors is large, this prevents the WSN from achieving much higher
energy savings.
Spatial correlation of the desired sensor signal components following a specific mathematical model is considered
in \cite{Chaudhary12}, where the noise components are again assumed to be uncorrelated.

% Star topology and complete signal models.
Spatial correlation of both the desired signal and noise components has been addressed more recently.
In \cite{DeLaHuchaArce16, DLHA17-01} a greedy bit depth allocation algorithm is proposed for linear MMSE estimation, but its convergence speed is slow when steps of one bit must be used.
Several works have considered a specific class of energy models which depend exponentially on the bit depths.
In \cite{Zhang18}, a bit depth allocation problem based on a linearly constrained minimum variance (LCMV) estimator is solved through convex relaxation of the minimum variance constraint, thereby loosing guarantees on the optimality of the solution.
This work has been extended to combine bit depth allocation with sensor selection under the same energy model in \cite{Zhang23}.
The convexity of bit depth allocation problems for linear MMSE estimation, where the bit depths are relaxed to real-valued variables instead of integer variables, is studied in \cite{DeLaHuchaArce20}. 
For the aforementioned exponential energy models, the minimization of energy consumption subject to an MMSE constraint can be transformed into a convex problem using a similar approach to \cite{Zhang18}, and can be solved efficiently through interior point methods \cite{BoydVandenberghe04}.
Finally, \cite{Amini19} presented a method  to minimize the MMSE subject to a constraint on communication bandwidth, achieving a similar performance with much lower complexity than a reference discrete optimization method.

% Multihop (tree) topology.
For a multihop WSN with a fusion centre planning the routing tree, \cite{Huang09, Huang11} present the solution to bit
depth allocations problems based on communication energy models depending both exponentially and linearly on the 
bit depths. 
However, each node averages the estimates from its neighbours instead of computing the optimal linear MMSE
estimator, and thus the obtained signal estimation performance is suboptimal.

% Energy models and their influence on bit depth allocation problems.
All of the the works concerned with energy minimization
\cite{GoldsmithDigComm06, Krasnopeev05, Huang09, Huang11, Zhang18, Zhang23, DeLaHuchaArce20} employ energy models based on communication channels corrupted by additive white Gaussian noise (AWGN) \cite{Goldsmith05, Huang09}, and assume an orthogonal multiple-access scheme.
The mathematical form of a communication energy model has important implications in the difficulty of the resulting
bit depth allocation problem.
Energy models depending exponentially on the sensor bit depths have been obtained under the additional assumption 
that the order of the QAM constellation scales with the sensor bit depth \cite{GoldsmithDigComm06}, and based on 
channel capacity \cite{Huang09, Huang11, Zhang18, Zhang23}. 
Together with the aforementioned assumptions, these works show how a set of bit depths close  to the optimal 
allocation can be obtained for these models through the solution of a single convex optimization problem. 
The conditions for the convexity of bit depth allocation problems for these energy models and linear MMSE estimation 
are given in \cite{DeLaHuchaArce20}. 
On the other hand, energy models depending linearly on the sensor bit depths are obtained when the modulation 
order is fixed, which is a fairly common configuration \cite{Goldsmith05, Rosas12}.
However, for these models bit depth allocation has only been studied under either a suboptimal averaging estimator
in multi-hop topologies \cite{Huang11} or  a suboptimal BLUE estimator with the additional assumption that all sensors are always active \cite{Krasnopeev05}.

% Contribution of this paper.
\subsection{Contribution}
In this paper we focus on bit depth allocation for linear MMSE signal estimation in a WSN with a star topology,
a communication energy model depending linearly on the sensor bit depths, and a general signal model that considers
correlated noise components across the different sensor nodes.
Even after relaxing the integer bit depths to real variables, this bit depth allocation problem cannot be transformed or
relaxed into a convex optimization problem. 
However, as we show, it can be expressed as a difference of convex (DC) problem \cite{LeThi2018_30Years} through two different approaches.
This fact allows to employ the concave-convex procedure (CCP), a majorization-minimization (MM) method to find local
solutions for non-convex problems that involve the difference of convex functions 
\cite{Yuille03, Lanckriet09, Lipp16, Shen16DCCP}.
CCP relies on the ability to solve convex problems efficiently in order to find a local optimum of the initial problem
through several iterations.
To do so, the concave terms are replaced by their linear approximations, which are global convex upper bounds,
and the solution of the resulting convex approximate problem becomes the initial point for the next iteration.

The approach we propose relies on the fact that the MMSE as a function of the quantization noise powers is concave, 
i.e., its negative is convex, while the total communication energy remains a convex function.
We show that the resulting convex approximate problem can be efficiently solved with a worst-case number of 
iterations equal to the number of sensors in the WSN, and where each iteration is computationally cheap.
This approach results in the main algorithm we propose, CCP-linMMSE.

% Structure of the paper.
The paper is structured as follows.
In Section \ref{sec:ProbForm} we formulate the bit depth allocation problem and motivate the use of CCP to solve it.
In Section \ref{sec:CCPLinearMMSE} we describe the CCP-linMMSE algorithm, we derive the solution to the convex
approximate problem that arises  at each iteration, and we analyze the convergence of CCP-linMMSE to a stationary
point of the original problem.
In Section \ref{sec:Simulations} we provide numerical simulations to support our analysis and compare CCP-linMMSE 
with other optimization algorithms.
Finally, in Section \ref{sec:Conclusions} we present our conclusions.

\section{Problem formulation} \label{sec:ProbForm}
\subsection{Signal model} \label{subsec:SignalModel}
% Signal assumptions
We consider a WSN composed of \(K\) sensors (channels), where the set of sensors is denoted by 
\(\mathcal{K} = \{ 1, \dots, K \} \).
The signal samples collected by the \(k\)-th sensor are denoted by \(y_k[t]\), where \(t \in \mathbb{N}\) denotes the
discrete time index.
For conciseness, we will omit the time index in the subsequent sections unless referring to a specific signal sample.
We assume that the signals \(y_k\) are complex-valued in order to allow frequency domain descriptions such as, e.g., 
the short-time Fourier transform (STFT).
Furthermore, all sensor signals are assumed to be realizations of short-term wide-sense stationary and ergodic 
stochastic processes.

% Signal model
The sensor signal \(y_{k}\) is modelled as the combination of a desired signal component \(x_{k}\) and an undesired 
zero-mean noise component \(v_{k}\), which can be expressed as 
\begin{align} \label{eq:SensorSignalModel}
	y_{k}  = x_{k} + v_{k}, \quad k \in \{1, \dots, K\} ,
\end{align} 
where we assume that \(E \{ x_k  v_m^*\} = 0\) \(\forall k, m \in \mathcal{K}\), where \(E\{\cdot\}\) denotes the
expectation operator and \((\cdot)^*\) denotes the complex conjugate.
The sensor signals can be stacked in the \(K \times 1\) vector \(\mathbf{y} = [y_1, \dots, y_K]^T\).
The \(K \times 1\) vectors \(\mathbf{x}\) and \(\mathbf{v}\) are defined in a similar way.
We remark that no statistical distribution, Gaussian or otherwise, is assumed on the sensor signals \(y_{k}\) or
their components \(x_{k}, v_{k}\).

\subsection{Linear MMSE estimation} \label{subsec:LMMSE}
% Centralized architecture
For the sake of simplicity, we assume a star topology for the WSN where all sensors send their signal samples to a fusion centre (FC), which then has access to the complete set of sensor signals \(\mathbf{y}\).

The goal of the WSN is to estimate a desired signal \(d\), which is correlated or equal to any one of the unobservable desired signal components \(x_k\), by applying the linear MMSE estimator to the sensor signals \(\mathbf{y}\).
A common example is where \(d\) is chosen as the desired signal in one of the sensors \cite{Cornelis11, DLHA17-01, Bertrand14-125}, i.e., \(d = x_k\) for some \(k\).
The linear MMSE estimator\footnote{Note that the linear MMSE estimator is optimal in the maximum likelihood sense if the noise is normally distributed. 
However, even if this is not the case, the linear MMSE estimator is still applicable and provides the lowest MSE that can be achieved with a linear estimator \cite{Kay93}.}
\(\hat{\mathbf{w}}\) is found by minimizing the mean squared error function, i.e.,
\begin{align} \label{eq:MMSEProblem}
	\hat{\mathbf{w}} = \underset{\mathbf{w}}{\text{arg min }} E \left \{ |d - \mathbf{w}^H \mathbf{y} |^2 \right \} \, ,
\end{align}
where \(( \cdot )^H\) denotes the conjugate transpose.
Assuming that the sensor signal correlation matrix \(\mathbf{R}_{yy} = E \left \{ \mathbf{y} \mathbf{y}^H \right \} \) 
has full rank\footnote{This assumption is typically satisfied in practice due to the presence of a 
noise component in each sensor that is independent of other sensor signals.}, the unique solution to 
(\ref{eq:MMSEProblem}) is given by
\begin{align} \label{eq:SolutionLMMSE} 
	\mathbf{\hat{w}} = \mathbf{R}_{y y}^{-1} \, \mathbf{r}_{yd} \, ,
\end{align}
where \(\mathbf{r}_{yd} = E \{ \mathbf{y} d^*\}\).
The estimate of the desired signal \(d\) is then given by
\begin{align} \label{eq:EstimateCentral}
	\hat{d} = \hat{\mathbf{w}}^H \mathbf{y} \, .
\end{align}
% MMSE no quantization.
The linear minimum mean squared error is given by
\begin{align} \label{eq:MMSE}
	 E \left \{ |d - \hat{\mathbf{w}}^H \mathbf{y} |^2 \right \} = 
	 P_d - \mathbf{r}_{yd}^H \mathbf{R}_{yy}^{-1} \mathbf{r}_{yd} \, .
\end{align}
where \(P_d = E\{|d|^2\}\) is the desired signal power.

% Estimation second order statistics.
Using the assumption of short-term ergodicity, the matrix \(\mathbf{R}_{yy}\)  can be estimated through sample 
averaging, e.g., using a sliding window or a forgetting factor.
In some applications, e.g., when working with pilot or training sequences, it is possible to estimate \(\mathbf{r}_{y d}\)
through sample averaging.
However, when the desired signal \(d\) is unobservable and unknown, sample averaging is impossible for 
\(\mathbf{r}_{y d}\), and hence its estimation must be done indirectly through strategies specific to each application.
For example, in cases where \(d = x_k\) and where the desired source signal has an `on-off' behaviour, as in speech 
\cite{Cornelis11, Markovich13, Hassani15-68} or EEG signal enhancement \cite{Bertrand14-125}, the noise correlation
matrix \(\mathbf{R}_{v v} = E \{ \mathbf{v} \mathbf{v}^H \}\) can be estimated during periods when the desired  source 
is inactive.
Together with the assumption that \(\mathbf{x}\) and \(\mathbf{v}\) are uncorrelated\footnote{For the sake of easy 
exposition we have also assumed that \(\mathbf{v}\) is zero-mean.}, 
this allows to obtain \(\mathbf{r}_{yd}\) as the \(k\)-th column of the desired signal covariance matrix
\(\mathbf{R}_{xx} = E \left \{ \mathbf{x} \mathbf{x}^H \right \}\), which can be estimated as 
\(\mathbf{R}_{xx} = \mathbf{R}_{yy} - \mathbf{R}_{vv}\) \cite{Cornelis11}.
More refined data-driven techniques to estimate \(\mathbf{r}_{yd}\) have been studied in 
\cite{SerizelGEVD, Hassani16}.

\subsection{Uniform quantization} \label{subsec:UnifQuant}
In practical applications of WSNs, the samples of each sensor signal \(y_k\) are quantized prior to their transmission
using a certain bit depth \(b_k\), which introduces a quantization error\footnote{Note that \(y_k\) could be 
already quantized, e.g., when it is acquired by the analog-to-digital converter of the \(k\)-th sensor. In that situation,
\(e_k\) would represent the error from changing the bit resolution of \(y_k\).} \(e_k\).
The uniform quantization of \(y_k\) with \(b_k\) bits can be expressed as
\begin{align} \label{eq:Quantization}
	Q(y_k) = \Delta_{b_k} \, \left ( \left \lfloor \frac{y_k}{\Delta_{b_k}} \right \rfloor + \frac{1}{2} \right ),
\end{align}
where \(\lfloor \cdot \rfloor\) denotes the floor function and \(\Delta_{b_k} = D_k/2^{b_k}\).
The parameter \(D_k\) is the dynamic range of the sensor signal \(y_k\), which is typically selected to be a few standard
deviations around its mean or \(2 \, \text{max}(|y_k|)\).
Since the dynamic range is specific to each sensor signal, it needs to be communicated to the FC.
The quantization error at the \(k\)-th sensor is then defined as
\begin{align} \label{eq:QuantError}
	e_k = Q(y_k) - y_k \, .
\end{align}
The mathematical properties of the quantization error \(e_k\) have been thoroughly studied 
\cite{Sripad77, Gray90, Gray98}. 
It has been shown that the quantization error is zero mean, and that \(y_k\) and \(e_k\) are 
uncorrelated under certain technical conditions on the characteristic function of \(y_k\) \cite{Sripad77}.
Under the same conditions, the power of the quantization error is given by
\begin{align} \label{eq:MSEQuantNoise}
	p_k = E\{|e_k|^2\} = \frac{\Delta_{b_k}^2}{12} = \frac{D_k^2}{12} \frac{1}{2^{2 b_k}}\, . 
\end{align}

Following the uniform quantization model, we can express the quantized sensor signal samples of the \(k\)-th sensor as 
\begin{align} \label{eq:QuantSensorSignals}
	y_{k,e} = y_k + e_k \, ,
\end{align}
which can then be stacked in the \(K \times 1\) vector 
\begin{align} \label{eq:QuantSignalsVector}
	\mathbf{y}_e = [y_{1,e}, \dots, y_{K,e}]^T = \mathbf{y} + \mathbf{e} \, ,
\end{align}
where \(\mathbf{e} = [e_1, \dots, e_K]^T\).
Using the statistical properties of \(e_k\) we can find the second-order statistics of the quantized sensor signals.
As \(e_k\) is assumed to be uncorrelated with \(d\), the cross-correlation remains unchanged, i.e.
\begin{align} \label{eq:CrossCorr_ye_d}
	\mathbf{r}_{y_e d} = E \left \{ (\mathbf{y} + \mathbf{e}) d^* \right \} = \mathbf{r}_{y d} \, .
\end{align}
The quantized sensor signal covariance matrix is given by
\begin{align} \label{eq:Ryyerror}
  \mathbf{R}_{y_e y_e} = E \left \{ ( \mathbf{y} + \mathbf{e} ) (\mathbf{y} + \mathbf{e} )^H \right \} =
  \mathbf{R}_{yy} + \mathbf{R}_{ee} \, ,
\end{align}
where \(\mathbf{R}_{ee}\) is the error correlation matrix, given by 
\begin{align} \label{eq:ReeDef} 
	\mathbf{R}_{ee} = E \left \{\mathbf{e} \mathbf{e}^H \right \} = \text{diag}(\mathbf{p}) \, ,
\end{align}
where \(\mathbf{p} = [p_1, \dots, p_K]^T\) is the \(K \times 1\) vector collecting the quantization error powers.
Note that (\ref{eq:Ryyerror}) and the diagonal structure of \(\mathbf{R}_{ee}\) in (\ref{eq:ReeDef}) follow from the 
assumptions that the quantization errors \(e_k\) are uncorrelated with every sensor signal \(y_m\) and every other
quantization error \(e_n\), \(\forall m, n \in \mathcal{K}, n \neq k\).

Using (\ref{eq:SolutionLMMSE}), (\ref{eq:CrossCorr_ye_d}) and (\ref{eq:Ryyerror}), we can find the optimal linear MMSE
estimator for the desired signal \(d\) using the quantized sensor signals \(\mathbf{y}_e\) as
\begin{align} \label{eq:MMSEQuantEstimator}
	\mathbf{w} (\mathbf{p}) = 
	\underset{\mathbf{w}}{\text{arg min }} E \left \{ |d - \mathbf{w}^H \mathbf{y}_e |^2 \right \} =
	\mathbf{R}_{y_e y_e}^{-1} \mathbf{r}_{yd} \, .
\end{align}
The MMSE corresponding to a specific set of quantization noise powers \(\mathbf{p}\), denoted by \(J(\mathbf{p})\), 
can be found from (\ref{eq:MMSE}) as
\begin{align} \label{eq:MMSE_Quant}
	 J(\mathbf{p}) = E \left \{ |d - \mathbf{w} (\mathbf{p})^H \mathbf{y}_e |^2 \right \} =  
	 P_d - \mathbf{r}_{yd}^H \mathbf{R}_{y_e y_e}^{-1} \mathbf{r}_{yd} \, ,
\end{align}
where  \(\mathbf{R}_{y_e y_e}\) depends on \(\mathbf{p}\) through (\ref{eq:Ryyerror}) and (\ref{eq:ReeDef}), and
\(\mathbf{p}\) depends on the bit depths \(\mathbf{b} = [b_1, \dots, b_K]^T\) through (\ref{eq:MSEQuantNoise}).

\subsection{Energy model} \label{subsec:energymodel}
As we explained in Section \ref{sec:Introduction}, most of the work in the literature has been focused on 
assumptions which lead to energy models depending exponentially on the bit depths \(b_k\).
Such exponential models are convenient from a mathematical perspective, as they can be cast to convex optimization
problems under certain conditions \cite{DeLaHuchaArce20}.
However, here we focus on an energy model depending linearly on the bit depths, which is a more realistic model for
orthogonal multiple-access schemes where the modulation order is fixed \cite{Goldsmith05, Rosas12}.
In particular, we focus on the energy model developed in \cite{Goldsmith05} for the transmission and reception of a
single packet between two nodes using M-QAM modulation on an additive white Gaussian noise (AWGN) channel.
This model also coincides with \cite{Rosas12} under the assumption that no retransmissions of the packet take place.
The treatment of retransmissions is outside of the scope of this paper.
We assume that all nodes have the same communication bandwidth \(W\) available, use the same order 
\(\mathcal{M}\) of QAM modulation, i.e. they use the same number of bits per symbol 
\(b_s = \text{log}_2 \mathcal{M}\),
and the WSN follows an orthogonal multiple-access scheme such as, e.g., time or  frequency division multiple access
(TDMA and FDMA respectively).
The energy \(E_k(b_k)\) spent in communicating to the FC \(L\) samples of the sensor signal \(y_k\), quantized 
using \(b_k\) bits, is given by
\begin{align} \label{eq:EnergyModel}
	E_k(b_k) = \left ( \sigma_k (2^{b_s} - 1) + \chi_k \right ) L \frac{b_k}{b_s} \, ,
\end{align}
where
\(\sigma_k = \frac{4}{3} N_{f,k} N_{0,k} G_{d,k} r_k^{\alpha} \frac{\xi_k}{\gamma_{\text{pa},k}} 
\text{ln} \left ( \frac{2}{p_{b,k}} \right ) \) and \(\chi_k = \frac{P_{c,k}}{W}\). 
These variables represent the receiver noise figure \(N_{f,k}\), the power spectral density of the AWGN \(N_{0,k}\), 
the bit error rate (BER) \(p_{b,k}\), the distance between the \(k\)-th node and the FC \(r_k\), the path-loss exponent
\(\alpha\), the transmitter gain \(G_{d,k}\), the power amplifier peak-to-average ratio \(\xi_k\), the drain efficiency
\(\gamma_{\text{pa},k}\), and the total (transmitter plus receiver) circuit power \(P_{c,k}\), excluding the power 
amplifier.
We can summarize the energy model (\ref{eq:EnergyModel}) as
\begin{align} \label{eq:EnergyModelSumm}
	E_k(b_k) = \rho_k b_k \, ,
\end{align}
which is linear in the bit depth \(b_k\), and where \(\rho_k = L b_s^{-1} \left ( \sigma_k (2^{b_s} - 1) + \chi_k \right ) \) 
is the energy per bit used to quantize the sensor signal \(y_k\).
Note that \(\rho_k\) is an a priori known node-dependent scalar that is independent of \(b_k\), and where 
\(\rho_k \geq 0, \, \forall k \in \mathcal{K}\).
We are not including the energy spent by the transceiver going from sleep to active mode since it is an additional
term independent of the number of transmitted bits and much smaller than the energy spent in active mode 
\cite{Goldsmith05}.

Naturally, the communication channels also introduce bit errors, characterized by the BER \(p_{b,k}\), which
increase the MMSE (\ref{eq:MMSE_Quant}) due to quantization.
Upper bounds on the total MMSE including channel bit errors have been derived in 
\cite{GoldsmithDigComm06, Chaudhary12} for their specific signal models under the key simplifying assumption that 
the estimator applied for ideal channels (i.e., zero BER) is also used in the case of non-ideal channels (i.e., non-zero 
BER).
Here, we assume that the BER is sufficiently low in the sense that the influence of quantization noise on the MMSE
is much larger than the effect of the BER on the MMSE.
Therefore, and due to the complexity of the general case, we consider the effect on the MMSE due to channel bit errors beyond the scope of this paper.

\subsection{Bit depth allocation} \label{subsec:BitDepthAlloc}
A bit depth allocation problem can be formulated as the minimization of the energy consumed in transmitting the 
sensor signals over the WSN while satisfying a performance requirement, i.e., the MMSE (\ref{eq:MMSE_Quant}) not 
exceeding a certain value.
Using the energy model (\ref{eq:EnergyModelSumm}) and the MMSE corresponding to a set of bit depths 
\(\mathbf{b}\) (\ref{eq:MMSE_Quant}), the problem can be expressed as
\begin{align} \label{eq:LinearBitAlloc}
	\underset{\mathbf{b}}{\text{min}} & \sum_{k=1}^K \rho_k b_k \nonumber \\
	\text{s.t.} \, 
	& P_d - \mathbf{r}_{yd}^H \left ( \mathbf{R}_{y y} + \mathbf{R}_{e e} \right )^{-1} \mathbf{r}_{yd} \leq \eta \, , \\
	& b_k \in \mathbb{Z}_+, \quad b_k \leq b_{\text{max}}, \forall k \nonumber ,
\end{align}
where \(\mathbb{Z}_+\) is the set of non-negative integers, \(b_{\text{max}}\) is the maximum bit depth allowed, 
\(\eta\) is a user-defined threshold on the allowed MSE, and \(\mathbf{R}_{ee}\) is the quantization error
covariance matrix given in (\ref{eq:ReeDef}).
As exhaustive integer search is impractical unless the problem is very small, we relax the bit depths to be real-valued
variables, i.e. \(b_k \in \mathbb{R}_+ \, \forall k\).
Any real solution then needs to be transformed back to \(\mathbb{Z}_+\), e.g., through rounding
as in \cite{Huang09, GoldsmithDigComm06, Huang11}, randomized rounding \cite{Zhang18} or a greedy heuristic
\cite{Chaudhary12}.
This discretization may lead to suboptimality, but from our numerical experiments presented in Section 
\ref{sec:Simulations} its impact appears to be minor.

It is readily seen that problem (\ref{eq:LinearBitAlloc}) is non-convex in the variable \(\mathbf{b}\) due to the MMSE
constraint.
As we show in the remainder of the paper, after the relaxation \(b_k \in \mathbb{R}_+ \, \forall k\), 
(\ref{eq:LinearBitAlloc}) can be expressed as a difference of convex (DC) problem {\cite{LeThi2018_30Years}, and thus we select CCP \cite{Yuille03, Lanckriet09, Lipp16, Shen16DCCP} as the basis for our proposed algorithm.
There are two main advantages of CCP over other optimization methods \cite{Lipp16}.
The first one is that the iterates in CCP retain more information than other popular methods, such as sequential 
quadratic programming (SQP) \cite{Nocedal06}, since only the concave terms are linearized while the full information 
of the convex terms is preserved.
The second one is that the upper bounds of the concave terms are global, and thus there is no need to limit the 
progress at each iteration to a region where the approximations are valid, like e.g. trust regions.

Our proposed method CCP-linMMSE, which we discuss in Section \ref{sec:CCPLinearMMSE}, relies on the MMSE 
function being a concave function of the quantization noise powers, as we show in Appendix \ref{app:ConcavityMMSE}.
This allows to use CCP by linearizing the MMSE function at each iteration, resulting in convex approximate problems 
with only linear constraints and whose solution can be found efficiently, as we show in Section
\ref{subsec:SolApproxCompl}.

\section{CCP algorithm based on linearization of the MMSE constraint} \label{sec:CCPLinearMMSE}
The first step is to express the cost function of (\ref{eq:LinearBitAlloc}) in the quantization noise power variable
\(\mathbf{p}\), whose elements are defined in (\ref{eq:MSEQuantNoise}), through the change of variables
\(b_k = 2^{-1} \left ( \delta_k - \text{log}_2 p_k \right ) \), where \(\delta_k = \text{log}_2 (12^{-1} D_k^2)\).
The energy model (\ref{eq:EnergyModelSumm}) becomes 
\begin{align} \label{eq:EnergyModelOneSensor}
	\tilde{E}_k (p_k) = E_k(2^{-1} \left ( \delta_k - \text{log}_2 p_k \right )),
\end{align}
 and thus the cost function of 
(\ref{eq:LinearBitAlloc}) is given by \(\tilde{E} (\mathbf{p}) = \sum_{k \in \mathcal{K}} \tilde{E}_k (p_k)\).
The problem (\ref{eq:LinearBitAlloc}) then becomes
\begin{align} \label{eq:LinearBitAlloc_2}
	\underset{\mathbf{p}}{\text{min}} & \sum_{k=1}^K \frac{1}{2} \rho_k \left ( \delta_k - \text{log}_2 p_k \right )
	\nonumber \\
	\text{s.t.} \, 
	& J(\mathbf{p}) \leq \eta \, , \\
	& p_{\text{min},k} \leq p_k \leq p_{\text{max},k}, \forall k \nonumber ,
\end{align}
where \(J(\mathbf{p})\) is the MMSE (\ref{eq:MMSE_Quant}), whose explicit expression as a function of 
\(\mathbf{p}\) is given by
\begin{align} \label{eq:MMSE_Function_QPower}
	J(\mathbf{p}) = P_d - \mathbf{r}_{yd}^H \left ( \mathbf{R}_{y y} + \text{diag}(\mathbf{p}) \right )^{-1} 
	\mathbf{r}_{yd} ,
\end{align}
\(p_{\text{max},k} = \frac{D_k^2}{12} = 2^{\delta_k}\) and
\(p_{\text{min},k} = 2^{\delta_k - 2 b_{\text{max}}}\), which correspond to \(b_k = 0\) and \(b_k = b_{\text{max}}\) 
respectively.
Once a solution \(\mathbf{p}^*\) to (\ref{eq:LinearBitAlloc_2}) is found, it must be transformed back to integer bit 
depths \(\mathbf{b}^* \in \mathbb{Z}_+\), as explained in Section \ref{subsec:BitDepthAlloc}.
The simplest method is rounding, i.e. 
\(b_k^* = \text{round} \left ( 2^{-1} \left ( \delta_k - \text{log}_2 p_k^* \right ) \right ) \).

\subsection{Formulation of the convex approximate problem} 
\label{subsec:ConvexApproxProb}
\begin{algorithm}[t]
\caption{CCP-linMMSE algorithm to solve the bit depth allocation problem (\ref{eq:LinearBitAlloc}).}
\label{alg:ccp_bitalloc}
\begin{algorithmic}[1]
	\STATE Initialize \(\mathbf{p}_0 \in \text{Dom } J\) to be a feasible point of (\ref{eq:LinearBitAlloc_2}),  
	\(\varepsilon > 0\).
	\STATE \(n := 0\)
	\REPEAT 
		\STATE \textit{Majorization:} 
		Compute \(J(\mathbf{p}_n), \nabla J(\mathbf{p}_n), 
		\bar{\eta}_n = \eta - J(\mathbf{p}_n) + \nabla J(\mathbf{p}_n)^T \mathbf{p}_n\).
		\STATE \textit{Minimization:} 
		Use Algorithm \ref{alg:search_dual} to find \(\lambda'_{n+1}\), and
		set \(\mathbf{p}_{n+1}\) to (\ref{eq:SolutionConvexApprox}).
		\STATE \textit{Update iteration} \(n := n+1\).
 	\UNTIL Convergence criterion is met, e.g. \(\tilde{E}(\mathbf{p}_n) - \tilde{E}(\mathbf{p}_{n+1}) \leq \varepsilon\).
 	\STATE Convert the solution \(\mathbf{p}^*\) back to integer bit depths, e.g. 
 	\(b_k^* = \text{round} \left ( 2^{-1} \left ( \delta_k - \text{log}_2 p_k^* \right ) \right ) \).
\end{algorithmic}
\end{algorithm}

Problem (\ref{eq:LinearBitAlloc_2}) remains non-convex due to the MMSE constraint.
In fact, the MMSE function \(J(\mathbf{p})\) is concave on its domain 
\(\text{Dom } J = \{ p_k \geq 0,  k \in \{1, \dots, K\} \} \), as we show in Appendix \ref{app:ConcavityMMSE}.
Due to the concavity of \(J (\mathbf{p}\)), its linearization around \(\mathbf{p}_n\) is a global upper bound 
\cite{BoydVandenberghe04}, i.e.,
\begin{align} \label{eq:LinearizedMMSE}
	J(\mathbf{p}) \leq J(\mathbf{p}_n) + \nabla J(\mathbf{p}_n)^T (\mathbf{p} - \mathbf{p}_n), 
	\forall \mathbf{p}, \mathbf{p}_n \in \text{Dom } J ,
\end{align} 
where \cite{DLHA17-01}
\begin{align} \label{eq:MMSEGradient}
	\nabla J(\mathbf{p}_n) = \left | \mathbf{w} (\mathbf{p}_n) \right |^2 ,
\end{align}
where \(| \cdot |^2\) is an element-wise operator, and 
\begin{align} \label{eq:MMSEGradExplicit}
	\mathbf{w} (\mathbf{p}_n) = \left ( \mathbf{R}_{y y} + \text{diag}(\mathbf{p}_n) \right )^{-1} \mathbf{r}_{yd} 
\end{align}
is the optimal linear MMSE estimator (\ref{eq:MMSEQuantEstimator}) at the current allocation of quantization noise
power \(\mathbf{p}_n\).
Note that the subscript \(n\) is used from now on as an iteration index.
By using the upper bound in (\ref{eq:LinearizedMMSE}), which is convex since it is a linear function, 
the problem (\ref{eq:LinearBitAlloc_2}) is replaced with the following convex approximate problem 
\begin{align} \label{eq:LinearBitAlloc_Linearized}
	\underset{\mathbf{p}}{\text{min}} & \sum_{k=1}^K \frac{1}{2} \rho_k 
	\left ( \delta_k - \text{log}_2 p_{k} \right ) \nonumber \\
	\text{s.t.} \, 
	 &\nabla J(\mathbf{p}_n)^T \mathbf{p} \leq \bar{\eta}_n \, , \\
	 & p_{\text{min},k} \leq p_k \leq p_{\text{max},k}, \forall k \nonumber ,
\end{align}
where \(\bar{\eta}_n = \eta - J(\mathbf{p}_n) + \nabla J(\mathbf{p}_n)^T \mathbf{p}_n\).
The solution to (\ref{eq:LinearBitAlloc_Linearized}), denoted by \(\mathbf{p}_{n+1}\), can be used to formulate a new
problem of the same form as (\ref{eq:LinearBitAlloc_Linearized}) by substituting \(\mathbf{p}_n\) with
\(\mathbf{p}_{n+1}\), as can be seen in Algorithm \ref{alg:ccp_bitalloc}.
The convergence of the sequence of approximate solutions \(\{\mathbf{p}_n\}\) and the required feasibility of the
initial point \(\mathbf{p}_0\) are discussed in Section \ref{subsec:ConvergenceBitAlloc}.

\subsection{Solution to the convex approximate problem} \label{subsec:SolApproxCompl}
Since the problem (\ref{eq:LinearBitAlloc_Linearized}) is convex and all of its constraint functions are affine,
the Karush-Kuhn-Tucker (KKT) conditions are necessary and sufficient for the global optimality of 
the minimizer \(\mathbf{p}_{n+1}\) \cite{BoydVandenberghe04, Nocedal06}.
These conditions are given by:
\begin{subequations}
\begin{align} 
	\label{eq:PF1} & 
	\nabla J(\mathbf{p}_n)^T \mathbf{p}_{n+1} \leq \bar{\eta}_n 
	\, , \\
	\label{eq:PF2} & p_{\text{min},k} \leq p_{n+1,k} \leq p_{\text{max},k}, \forall k, \\
	\label{eq:DF_Comp} & \lambda_{n+1} \geq 0, \, \mu_{n+1,k} \geq 0, \, \nu_{n+1,k} \geq 0, \forall k, \\
	\label{eq:CS1} & \lambda_{n+1} 
	\left (\nabla J(\mathbf{p}_n)^T \mathbf{p}_{n+1} - \bar{\eta}_n 
	\right ) = 0 \, , \\
	\label{eq:CS2} & \mu_{n+1,k} (p_{\text{min},k} - p_{n+1,k}) = 0, \forall k, \\
	\label{eq:CS3} & \nu_{n+1,k} (p_{n+1,k} - p_{\text{max},k}) = 0, \forall k, \\
	\label{eq:LS_Comp} & \nabla \tilde{E}(\mathbf{p}_{n+1}) + \lambda_{n+1} \nabla J(\mathbf{p}_n) 
	- \boldsymbol{\mu}_{n+1} + \boldsymbol{\nu}_{n+1} = 0 \, ,
\end{align}
\end{subequations}
where \(\lambda_{n+1}, \mu_{n+1,k}, \nu_{n+1,k}, \forall k \in \mathcal{K}\), denote the optimal dual variables,
and where the latter two are stacked in the vectors \(\boldsymbol{\mu}_{n+1}\) and \(\boldsymbol{\nu}_{n+1}\)
respectively.

It is easy to verify from (\ref{eq:PF2}), (\ref{eq:DF_Comp}), (\ref{eq:CS2}) and (\ref{eq:CS3}) that \(\mu_{n+1,k} = 0\) if 
\(p_{n+1,k} > p_{\text{min},k}\), \(\mu_{n+1,k} \geq 0\) if \(p_{n+1,k} = p_{\text{min},k}\),  \(\nu_{n+1,k} = 0\) if 
\(p_{n+1,k} < p_{\text{max},k}\), and \(\nu_{n+1,k} \geq 0\) if \(p_{n+1,k} = p_{\text{max},k}\).
Note that \(\mu_{n+1,k} > 0\) and \(\nu_{n+1,k} > 0\) cannot be true at the same time for the same \(k\), as
\(p_{\text{max},k} \neq p_{\text{min},k}\).

Therefore, we divide the set of sensor indices \(\mathcal{K}\) in three disjoint sets, 
\(\mathcal{K}_{\mu} = \{ k \in \mathcal{K}: p_{n+1,k} = p_{\text{min},k} \}\), 
\(\mathcal{K}_{\nu} = \{ k \in \mathcal{K}: p_{n+1,k} = p_{\text{max},k} \}\)
and \(\mathcal{K}_{I} = \{k \in \mathcal{K}: p_{\text{min},k} < p_{n+1,k} < p_{\text{max},k}\}\).
In terms of the bit depth in (\ref{eq:LinearBitAlloc}), \(\mathcal{K}_{\mu}\) contains the sensors which will be assigned 
the maximum bit depth \(b_\text{max}\), \(\mathcal{K}_{\nu}\) the sensors which will be assigned 0 bits, i.e., 
do not transmit their sensor signal samples, and \(\mathcal{K}_I\) contains the sensors assigned any other bit depth.
Further on, we propose an algorithm to efficiently determine which nodes belong to \(\mathcal{K}_{\mu}\),
\(\mathcal{K}_{\nu}\) and \(\mathcal{K}_{I}\), respectively (see Appendix \ref{app:SearchDual}, which results in
Algorithm \ref{alg:search_dual}). 
If \(\mathcal{K}_{I}\) turns out to be empty, this means that \(\mathcal{K}_{\mu} \cup \mathcal{K}_{\nu} = \mathcal{K}\),
with one of them possibly being empty.
In this case, the solution of (\ref{eq:LinearBitAlloc_Linearized}) is straightforwardly found.
Therefore, we first focus on the case there \(\mathcal{K}_{I}\) is non-empty.

From (\ref{eq:CS1}) we know that \(\lambda_{n+1} = 0\), or
\(\nabla J(\mathbf{p}_n)^T \mathbf{p}_{n+1} = \bar{\eta}_n\), or both statements are true simultaneously.
However, if  \(\lambda_{n+1} = 0\) then (\ref{eq:LS_Comp}) would imply that 
\(\nabla \tilde{E}(\mathbf{p}_{n+1})_k = 0, \forall k \in \mathcal{K}_{I}\).
If \(\mathcal{K}_{I}\) is non-empty, this is impossible since the components of 
\(\nabla \tilde{E}(\mathbf{p})\) never vanish, as they are given by
\begin{align} \label{eq:GradEnergy}
	\nabla \tilde{E}(\mathbf{p})_k = \frac{- \rho_k}{2 \text{ln} 2} p_k^{-1} \, , \forall k \in \mathcal{K} .
\end{align}
We can then conclude that, if \(\mathcal{K}_I\) is non-empty, \(\lambda_{n+1} > 0\) and 
\begin{align} \label{eq:MMSELin_Equality}
	\nabla J(\mathbf{p}_n)^T \mathbf{p}_{n+1} = \bar{\eta}_n 
	\, .
\end{align}

From (\ref{eq:LS_Comp}), (\ref{eq:GradEnergy}) and (\ref{eq:MMSEGradient}) we can establish that
\begin{align} 
	\label{eq:MuLS} 
	& \frac{- \rho_k}{2 \text{ln} 2} \frac{1}{p_{\text{min},k}} + \lambda_{n+1} \left | w_k(\mathbf{p}_n) \right |^2 
	= \mu_{n+1,k},
	& \quad k \in \mathcal{K}_{\mu}, \\
	\label{eq:NuLS} 
	 & \frac{\rho_k}{2 \text{ln} 2} \frac{1}{p_{\text{max},k}} - \lambda_{n+1} \left | w_k(\mathbf{p}_n) \right |^2 = 
	 \nu_{n+1,k},
	&\quad k \in \mathcal{K}_{\nu}, \\
	\label{eq:I_LS} 
	& \frac{- \rho_k}{2 \text{ln} 2} \frac{1}{p_{n+1,k}} + \lambda_{n+1} \left | w_k(\mathbf{p}_n) \right |^2 = 0,
	& \quad k \in \mathcal{K}_{I} ,
\end{align}
where \(w_k(\mathbf{p}_n)\) is the \(k\)-th element of \(\mathbf{w} (\mathbf{p}_n)\) (\ref{eq:MMSEGradExplicit}).
For simplicity, we define \(\lambda'_{n+1} = 2 \text{ln} 2 \, \lambda_{n+1}\).
Since \(\mu_{n+1,k} \geq 0\) for \(k \in \mathcal{K}_{\mu}\) and \(\nu_{n+1,k} \geq 0\) for \(k \in \mathcal{K}_{\nu}\), 
we obtain the following conditions for \(\lambda'_{n+1}\) from (\ref{eq:MuLS}) and (\ref{eq:NuLS})
\begin{align} 
	\label{eq:Lambda_Cond1} 
	& \lambda'_{n+1} \geq \frac{\rho_k}{\left | w_k(\mathbf{p}_n) \right |^2 p_{\text{min},k}} 
	= \lambda_{\mu k}, & k \in \mathcal{K}_{\mu}, \\
	\label{eq:Lambda_Cond2} 
	 & \lambda'_{n+1} \leq \frac{\rho_k}{\left | w_k(\mathbf{p}_n) \right |^2 p_{\text{max},k}} =
	 \lambda_{\nu k}, & k \in \mathcal{K}_{\nu}. \\
	 \label{eq:Lambda_Cond3} 
	 & \lambda_{\nu k} < \lambda'_{n+1} < \lambda_{\mu k}, & k \in \mathcal{K}_I .
\end{align}
Using (\ref{eq:I_LS}) - (\ref{eq:Lambda_Cond3}) and the fact that 
\(p_{n+1,k} = p_{\text{min},k} \text{ if } k \in \mathcal{K}_{\mu}\)
and \(p_{n+1,k} = p_{\text{max},k} \text{ if } k \in \mathcal{K}_{\nu}\), we can express the optimal solution \(p_{n+1,k}\) 
as
\begin{subnumcases}{\label{eq:SolutionConvexApprox} p_{n+1,k} = } 
	p_{\text{max},k}
	& \( \text{if } \lambda'_{n+1} \leq \lambda_{\nu k} \) \label{eq:Pk_1} 
	\\ 
	\frac{\rho_k}{\left | w_k(\mathbf{p}_n) \right |^2} \frac{1}{\lambda'_{n+1}} 
	& \( \text{if } \lambda_{\nu k} < \lambda'_{n+1} < \lambda_{\mu k} \qquad \)  \label{eq:Pk_2}
	\\
	p_{\text{min},k} 
	& \( \text{if } \lambda'_{n+1} \geq \lambda_{\mu k} \) \label{eq:Pk_3}
\end{subnumcases}
All that remains is to find \(\lambda'_{n+1}\) through (\ref{eq:MMSELin_Equality}), i.e.,
\begin{align} 
	\nabla J(\mathbf{p}_n)^T \mathbf{p}_{n+1} =
	\sum_{k = 1}^K  \left | w_k(\mathbf{p}_n) \right |^2 p_{n+1,k} = 
	\bar{\eta}_n 	\, ,
\end{align}
which can be rewritten using (\ref{eq:Pk_1}), (\ref{eq:Pk_2}), (\ref{eq:Pk_3}) as
\begin{align} \label{eq:SumMMSEThr}
	& \sum_{k \in \mathcal{K}_{I} } \frac{\rho_k}{\lambda'_{n+1}} +
	\sum_{k \in \mathcal{K}_{\nu} }  \left | w_k(\mathbf{p}_n) \right |^2 p_{\text{max},k} + 
	\sum_{k \in \mathcal{K}_{\mu} }  \left | w_k(\mathbf{p}_n) \right |^2 p_{\text{min},k} \nonumber \\
	& = 
	\bar{\eta}_n 	\, .
\end{align}
This allows to compute \(\lambda'_{n+1}\)  as 
\begin{align} \label{eq:LambdaExpr}
	\lambda'_{n+1} = \frac{\rho_I}{\eta_I} \, ,
\end{align}
where \(\rho_I = \sum_{k \in \mathcal{K}_I} \rho_k\), 
\(\eta_I = \bar{\eta}_n - \eta_{\nu} - \eta_{\mu}\), 
\(\eta_{\nu} = \sum_{k \in \mathcal{K}_{\nu} }  \left | w_k(\mathbf{p}_n) \right |^2 p_{\text{max},k}\) and
\(\eta_{\mu} = \sum_{k \in \mathcal{K}_{\mu} }  \left | w_k(\mathbf{p}_n) \right |^2 p_{\text{min},k}\).
Therefore, we need to establish which sensor indices belong to \(\mathcal{K}_{\nu}\) and \(\mathcal{K}_{\mu}\)
in order to compute \(\lambda'_{n+1}\) from (\ref{eq:LambdaExpr}).
This can be achieved with a worst-case complexity of \(K\) iterations (under mild conditions)
through Algorithm \ref{alg:search_dual}, as explained in detail in Appendix \ref{app:SearchDual}.
Note that this algorithm also identifies the special cases where \(\mathcal{K}_{I}\) would be empty, which will 
result in the exclusive activation of the cases (\ref{eq:Pk_1}) and (\ref{eq:Pk_3}).
\begin{algorithm}[t]
\caption{Search algorithm for the optimal dual variable \(\lambda'_{n+1}\) of problem 
(\ref{eq:LinearBitAlloc_Linearized}), leading to the solution \(\mathbf{p}_{n+1}\)}
\label{alg:search_dual}
\begin{algorithmic}[1]
	\STATE Initialize \(\mathcal{K}_I = \mathcal{K}\), \(\mathcal{K}_{\nu} = \emptyset\), 
	\(\mathcal{K}_{\mu} = \emptyset\).
	\STATE Compute \(\lambda'_{n+1}\) through (\ref{eq:LambdaExpr}), i.e. 
	 \(\lambda'_{n+1} = \frac{\rho_I}{\bar{\eta}_n}\).
	where here \(\rho_I = \sum_{k = 1}^K \rho_k\).
	\IF{\(\lambda_{n+1}' < \underset{k}{\text{min}} \, \lambda_{\mu k}\)}
		\WHILE{\(\lambda'_{n+1}\) does not satisfy (\ref{eq:Lambda_Cond2}) and (\ref{eq:Lambda_Cond3})}
			\STATE \(\mathcal{K}_{\nu}^{\text{new}} = \{k \in \mathcal{K}_I :  \lambda'_{n+1} < \lambda_{\nu k}\} \)
			\STATE \( \mathcal{K}_{\nu}  \leftarrow  \mathcal{K}_{\nu} \cup \mathcal{K}_{\nu}^{\text{new}}\)
			\STATE \( \mathcal{K}_I  \leftarrow  \mathcal{K}_I \setminus \mathcal{K}_{\nu}^{\text{new}}\)
			\STATE Assign to \(\lambda'_{n+1}\) the result of (\ref{eq:LambdaExpr}).
		\ENDWHILE
	\ELSIF{\(\lambda_{n+1}' > \underset{k}{\text{max}} \, \lambda_{\nu k}\)}
		\WHILE{\(\lambda'_{n+1}\) does not satisfy (\ref{eq:Lambda_Cond1}) and (\ref{eq:Lambda_Cond3})}
			\STATE \(\mathcal{K}_{\mu}^{\text{new}} = \{k \in \mathcal{K}_I :  \lambda'_{n+1} > \lambda_{\mu k}\} \)
			\STATE \( \mathcal{K}_{\mu}  \leftarrow  \mathcal{K}_{\mu} \cup \mathcal{K}_{\mu}^{\text{new}}\)
			\STATE \( \mathcal{K}_I  \leftarrow  \mathcal{K}_I \setminus \mathcal{K}_{\mu}^{\text{new}}\)
			\STATE Assign to \(\lambda'_{n+1}\) the result of (\ref{eq:LambdaExpr}).
		\ENDWHILE
	\ENDIF
 	\STATE The optimal \(\lambda'_{n+1}\) is found, compute \(\mathbf{p}_{n+1}\) through 
 	(\ref{eq:SolutionConvexApprox}).
\end{algorithmic}
\end{algorithm}

\subsection{Convergence of the CCP bit depth allocation algorithm}
\label{subsec:ConvergenceBitAlloc}
We first show the feasibility of the iterates \(\{\mathbf{p}_n, n \in \mathbb{N}\}\) and the non-increasing  property of 
the cost function values at the iterates \(\{\tilde{E}(\mathbf{p}_n), n \in \mathbb{N}\}\).
Both of these properties are true in general for the CCP algorithm \cite{Yuille03, Lanckriet09, Lipp16}, and are included 
here for the convenience of the reader.
We will then use a result from \cite{Lanckriet09} to establish convergence to a stationary point.

If \(\mathbf{p}_n\) is a feasible point of the original bit depth allocation problem (\ref{eq:LinearBitAlloc_2}) then 
it is also a feasible point for the convex approximate problem (\ref{eq:LinearBitAlloc_Linearized}), 
as \(\bar{\eta}_n\) is non-negative in the first constraint of (\ref{eq:LinearBitAlloc_Linearized}).
The solution to (\ref{eq:LinearBitAlloc_Linearized}), i.e., \(\mathbf{p}_{n+1}\), is also a feasible point of 
(\ref{eq:LinearBitAlloc_2}).
To show this, we note that \(\mathbf{p}_{n+1}\), and in fact any feasible point of (\ref{eq:LinearBitAlloc_Linearized}),
satisfies the first constraint of (\ref{eq:LinearBitAlloc_Linearized}), for which the following inequalities can be 
obtained:
\begin{align}
	J(\mathbf{p}_{n+1}) \leq J(\mathbf{p}_n) + \nabla J(\mathbf{p}_n)^T (\mathbf{p}_{n+1} - \mathbf{p}_n) \leq \eta \, .
\end{align}
The right-most inequality follows from expanding the first constraint of (\ref{eq:LinearBitAlloc_Linearized}) using the 
definition of \(\bar{\eta}_n\), and the left-most inequality follows from the concavity of \(J(\mathbf{p})\) expressed in
(\ref{eq:LinearizedMMSE}).
It can be readily seen that \(J(\mathbf{p}_{n+1}) \leq \eta\), and therefore \(\mathbf{p}_{n+1}\) is a feasible point of 
(\ref{eq:LinearBitAlloc_2}).
Thus, if \(\mathbf{p}_0\) is chosen to be feasible, all iterates \(\{\mathbf{p}_n, n \in \mathbb{N}\}\) are feasible points 
of the original problem (\ref{eq:LinearBitAlloc_2}).
An easy choice for a feasible \(\mathbf{p}_0\) is the vector 
\(\mathbf{p}_{\text{min}} = [p_{\text{min},1}, \dots, p_{\text{min},K}]^T\), which also serves as a test for the feasibility
of (\ref{eq:LinearBitAlloc_2}) in general, since \(J(\mathbf{p}_{\text{min}}) > \eta\) would imply that the problem is
infeasible.

The evaluation of the cost function at the iterates satisfies the monotonic descent property
\(\tilde{E}(\mathbf{p}_{n+1}) \leq \tilde{E}(\mathbf{p}_n)\), since the function \(\tilde{E}(\mathbf{p})\) is minimized at
each iteration and the value \(\tilde{E}(\mathbf{p}_n)\) was already  achieved at \(\mathbf{p}_n\).
In fact, the cost function \(\tilde{E}(\mathbf{p})\) is strictly convex, since it is the non-negative weighted sum of strictly
convex functions due to \(- \text{log}_2 p_k\) being a strictly convex function\footnote{This can be easily shown since 
its second derivative is strictly positive on its domain.}.
As a consequence, the descent property is strict, i.e. if \(\mathbf{p}_{n+1} \neq \mathbf{p}_n\) then
\(\tilde{E}(\mathbf{p}_{n+1}) < \tilde{E}(\mathbf{p}_n)\).
The sequence \(\{\tilde{E}(\mathbf{p}_n), n \in \mathbb{N}\}\) is therefore strictly decreasing and, as it is bounded in 
the feasible set of problem (\ref{eq:LinearBitAlloc_2}), it must converge.
Note that this fact alone does not imply the convergence of \(\{\mathbf{p}_n, n \in \mathbb{N}\}\) to a single point,
nor is it sufficient to show that the convergence point is a stationary point of (\ref{eq:LinearBitAlloc_2}).

The convergence behaviour of the CCP algorithm is studied in depth in \cite{Lanckriet09}.
In particular, \cite[Theorem 10]{Lanckriet09} proves that, under certain conditions, all the limit points of any sequence 
of iterates generated by the CCP algorithm are stationary points of the original optimization problem.
The conditions of \cite[Theorem 10]{Lanckriet09} require both the cost and constraint functions involved in 
the original optimization problem to be differentiable,
the gradient of the concave part of the cost function to be continuous, and the feasible set to be compact.
It can be readily seen that all of these conditions are satisfied by problem (\ref{eq:LinearBitAlloc_2}).

The stationary points are those where necessary optimality conditions hold, i.e., the KKT conditions of the original
problem, provided that a suitable constraint qualification is satisfied.
The KKT conditions of problem (\ref{eq:LinearBitAlloc_2}) are given by
\begin{subequations}
\begin{align} 
	\label{eq:PFTr1} & P_d - 
	\mathbf{r}_{yd}^H \left ( \mathbf{R}_{y y} + \text{diag}(\mathbf{p}^*) \right )^{-1} \mathbf{r}_{yd} \leq \eta \, , \\ 
	\label{eq:PFTr2} & p_{\text{min},k} \leq p_k^* \leq p_{\text{max},k}, \forall k, \\
	\label{eq:DFTr} & \lambda^* \geq 0, \, \mu_k^* \geq 0, \, \nu_k^* \geq 0, \forall k, \\
	\label{eq:CSTr1} & \lambda^* 
	\left (P_d - \mathbf{r}_{yd}^H \left ( \mathbf{R}_{yy} + \text{diag}(\mathbf{p}^*) \right )^{-1} \mathbf{r}_{yd} - \eta 
	\right) = 0 \, , \\
	\label{eq:CSTr2} & \mu_k^* (p_{\text{min},k} - p_k^*) = 0, \forall k, \\
	\label{eq:CSTr3} & \nu_k^* (p_k^* - p_{\text{max},k}) = 0, \forall k, \\
	\label{eq:LS_Tr} & \nabla \tilde{E}(\mathbf{p}^*) + \lambda^* \nabla J(\mathbf{p}^*) 
	- \boldsymbol{\mu}^* + \boldsymbol{\nu}^* = 0 \, ,
\end{align}
\end{subequations}
where \(\nabla J(\mathbf{p}^*) =  \left | \mathbf{w} (\mathbf{p}^*) \right |^2 = 
\left | ( \mathbf{R}_{y y} + \text{diag}(\mathbf{p}^*) )^{-1} \mathbf{r}_{yd} \right |^2\), as seen from 
(\ref{eq:MMSEGradient}) and (\ref{eq:MMSEGradExplicit}).
In order for (\ref{eq:PFTr1})-(\ref{eq:LS_Tr}) to be neccesary conditions for a stationary point, also the linear
independence constraint qualification (LICQ) should be satisfied \cite{Nocedal06}.
In Appendix \ref{app:LICQ} we show that the LICQ is satisfied in general, except for some contrived cases which are 
not expected to emerge in practical applications.
The appendix also demonstrates that, if desired, the satisfaction of the LICQ at a certain point can be easily checked.

Note that, under the conditions of \cite[Theorem 10]{Lanckriet09}, oscillations of the sequence of iterates 
between stationary points where the same value of the cost function is achieved are still possible.
Nevertheless, if the cost function is the difference of two strictly convex functions, \cite{Lanckriet09} explains how to
formulate a stronger version of \cite[Theorem 10]{Lanckriet09}.
As we mentioned before, the cost function \(\tilde{E}(\mathbf{p})\) is a strictly convex function.
Therefore, this stronger version of \cite[Theorem 10]{Lanckriet09} guarantees that 
\( \| \mathbf{p}_{n+1} - \mathbf{p}_n \|_2 \to 0\) as \(n \to \infty\), and that either the sequence of
iterates \(\{\mathbf{p}_n, n \in \mathbb{N}\}\) converges or its set of limit points is a connected and compact subset of
the set of stationary points with the same cost function value \(\tilde{E}^*\).
If this set is finite, then any sequence of iterates generated by the CCP algorithm converges to some stationary point
in this set.
Note that, while both  \cite[Theorem 10]{Lanckriet09} and its stronger version are valid for any feasible initial point, 
the specific stationary point may indeed depend on the chosen initial point.

While the previous discussion ensures the convergence of the CCP-linMMSE algorithm, detailed in Algorithm 
\ref{alg:ccp_bitalloc}, to a stationary point of (\ref{eq:LinearBitAlloc_2}), the rate of convergence is still an open and challenging problem, as shown in \cite{Lanckriet09} \cite{LeThi2024} and references therein.
Results have been derived for specific classes of DC problems, such as in \cite{Yen12}, where at least linear convergence is proven for non-smooth cost functions with piecewise linear components on polyhedral feasible sets. 
For subanalytic objective functions satisfying certain smoothness conditions, \cite{LeThi2018} proves that the convergence rate depends on their Łojasiewicz exponent, whose computation is in general a challenging problem itself.

\section{Simulations} \label{sec:Simulations}
We study several WSNs in a two dimensional \(5 \times 5\) m area where the positions of the desired source, 
noise sources and sensors are randomly generated, and their number can vary across scenarios, as explained later on.
All source signals consist of \(10^5\) samples of complex Gaussian white stochastic processes with zero mean and unit
variance.
We use a narrow-band propagation model at a frequency \(f = 1\) kHz, where each source is observed at each sensor with a power attenuation factor inversely proportional to the distance between the corresponding source and the sensor, and a phase delay generated from a uniform distribution in \([- \pi, \pi]\).
Besides, independent AWGN is also present in each sensor signal, with zero mean and a variance of \(10^{-3}\) times the 
power of the corresponding sensor signal.
The second-order statistics \(\mathbf{R}_{yy}\) and \(\mathbf{r}_{yd}\) are estimated by averaging over all the signal
samples.
As we are interested in observing the effects of quantization, we compute \(\mathbf{r}_{yd}\) assuming knowledge of 
the desired signal \(d\) (the desired source as observed in the sensor with the highest SNR) to minimize estimation
errors.
For simplicity we choose \(\rho_k = r_k^2\), where \(r_k\) is the distance between the \(k\)-th sensor and the FC.

The proposed CCP-linMMSE method, detailed in Algorithm~\ref{alg:ccp_bitalloc}, is compared against three 
alternative methods to solve the bit depth allocation problem (\ref{eq:LinearBitAlloc}):
\begin{itemize}
	\item CCP-LMI, detailed in Algorithm \ref{alg:ccp_bitalloc_lmi}  in Appendix \ref{app:ccp_lmi}, which is also 
	based on CCP but uses a strategy from \cite{DeLaHuchaArce20} based on a linear matrix inequality (LMI)
	formulation of the MMSE constraint.
	Each convex sub-problem (\ref{eq:ccp_subprob_lmi}) is solved using the CVX package \cite{cvx, Grant08}.
	\item Sequential Quadratic Programming (SQP), a standard method for non-linear optimization \cite{Nocedal06}.
	We use the MATLAB function \textit{fmincon} with the default configuration to solve problem 
	(\ref{eq:LinearBitAlloc_2}), and then transform the solution to integer bit depths 
	using rounding, as it is the the simplest method, in the same way as the two previous algorithms.
	\item The greedy quantization method described in \cite{DeLaHuchaArce16, DLHA17-01}, which we refer to in 
	the following as GQM.
\end{itemize}
The initial point is always chosen to be \(\mathbf{p}_0 = \mathbf{p}_{\text{min}}\) for Algorithm \ref{alg:ccp_bitalloc},
\(\mathbf{u}_0 = 2^{2 b_{\text{max}}} \mathbf{1}_K\) for Algorithm \ref{alg:ccp_bitalloc_lmi}, and 
\(\mathbf{b}_0 = b_{\text{max}} \mathbf{1}_K\) for the greedy quantization method, where \(\mathbf{1}_K\) is
the \(K \times 1\) vector whose entries are all one.
Algorithm \ref{alg:ccp_bitalloc} and Algorithm \ref{alg:ccp_bitalloc_lmi} stop when 
\(\frac{E_n - E_{n+1}}{E_n} \leq 10^{-3}\), where \(E_n\) refers to \(\tilde{E}(\mathbf{p}_n)\) and \(\bar{E}(\mathbf{u}_n)\) 
respectively.
 
\subsection{Global and local convergence}
 \begin{figure}[!t]
    \centering
    \begin{subfigure}[t]{0.5\textwidth}
    % \begin{subfigure}[t]{0.85\textwidth} % Single column
        \includegraphics[width=8.5cm, height=5.5cm]{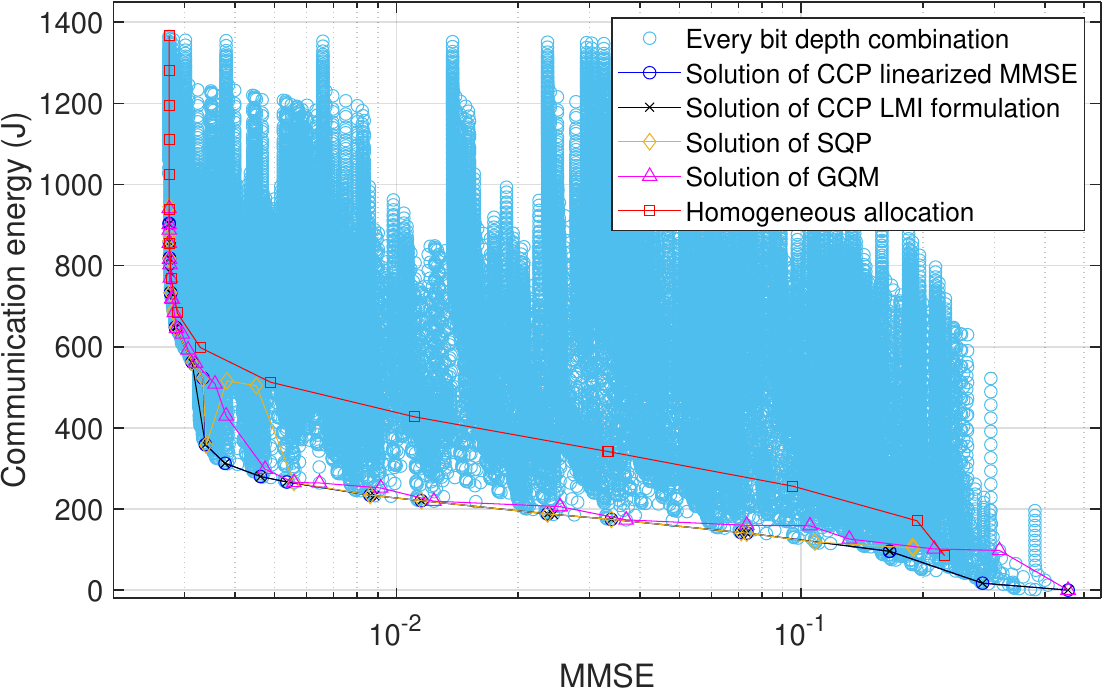}
        \caption{Scenario 1: The coordinates of the desired source are [4.64, 2.01] and of the noise sources 
        [1.89, 3.94]; [3.17, 1.40].}
        \label{fig:cloud_global}
    \end{subfigure}
    ~ %add desired spacing between images, e. g. ~, \quad, \qquad, \hfill etc. 
      % (!!) Change subfigure size when changing between single or double column.
    \begin{subfigure}[t]{0.5\textwidth}
    % \begin{subfigure}[t]{0.85\textwidth}
    	\includegraphics[width=8.5cm, height=5.5cm]{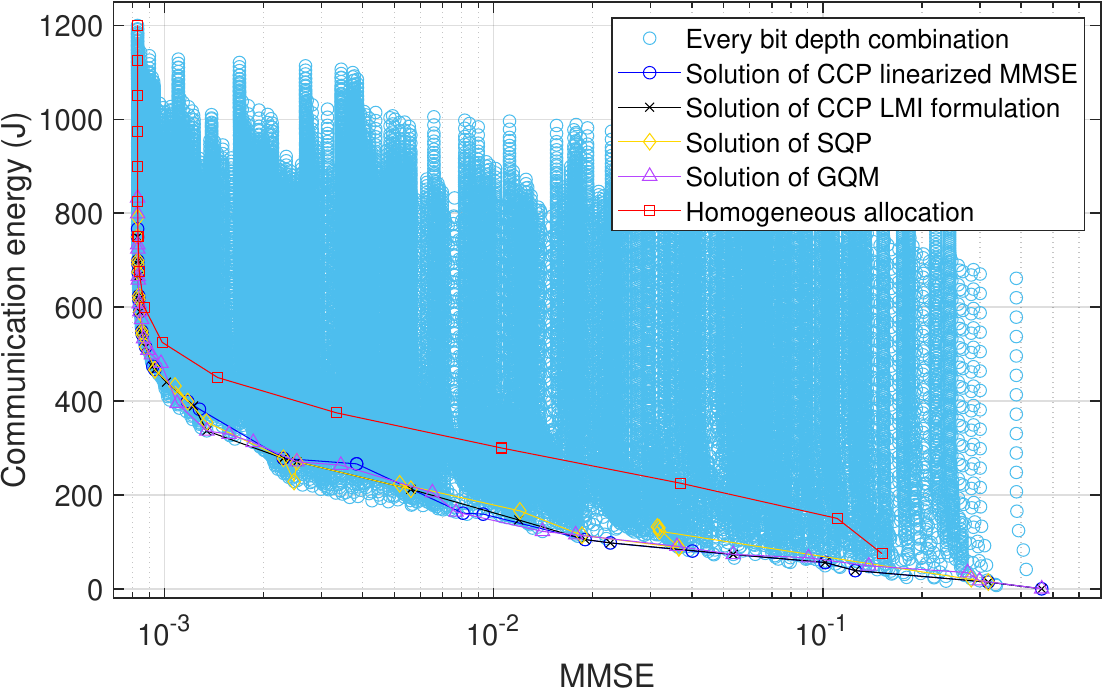}
        \caption{Scenario 2: The coordinates of the desired source are [4.24, 2.14] and of the noise sources 
        [4.47, 3.14]; [4.25, 3.94].}
        \label{fig:cloud_notglobal}
    \end{subfigure}
    \caption{Energy and MMSE corresponding to every bit depth combination, the solutions from Algorithm 
    \ref{alg:ccp_bitalloc} (CCP linearized MMSE), Algorithm \ref{alg:ccp_bitalloc_lmi} (CCP-LMI), SQP, GQM (greedy), 
     and homogeneous allocation (same bit depth for every sensor) for \(K = 4\) sensors.}
\end{figure}
 
Figures \ref{fig:cloud_global} and \ref{fig:cloud_notglobal} depict, for two different scenarios with \(K = 4\) sensors 
and two noise sources, the resulting total energy and MMSE from the bit depth allocation (\ref{eq:LinearBitAlloc}) for
different thresholds \(\eta\), solved through our proposed method CCP-linMMSE, detailed in Algorithm
\ref{alg:ccp_bitalloc}, through the CCP-LMI method detailed in Algorithm \ref{alg:ccp_bitalloc_lmi}, through SQP,
and through GQM.
For reference, the results from every bit depth combination from \(b_{\text{max}} = 16\) 
to 0 bits and from homogeneous allocation, i.e., where every sensor uses the same bit depth, are also depicted.
In Figure \ref{fig:cloud_global}, we can observe that the SQP and GQM methods sometimes get stuck in heavily 
suboptimal minima, whereas the solutions from Algorithm \ref{alg:ccp_bitalloc} come very close to the global optimal 
points for all values of \(\eta\) (in the plot, these are represented in the MMSE axis).
The small deviations are due to (a) the discretization of the solution, and (b) some possible mismatch with the 
quantization model used in (\ref{eq:ReeDef}).
However, in Figure \ref{fig:cloud_notglobal} we can clearly appreciate that for some values of \(\eta\) the convergence
point is not the global optimum, confirming that CCP-linMMSE will sometimes converge to a local minimum, as discussed in Section \ref{subsec:ConvergenceBitAlloc}.
In both figures, the behaviour of the CCP-LMI method is quite similar to that of the CCP-linMMSE method.
While both methods are based on the CCP, this is not necessarily guaranteed from a theoretical standpoint.
Nevertheless, as we shall see in Section \ref{subsec:CompTime_CCP}, CCP-linMMSE is significantly more
computationally efficient, making it more suited for adaptive and low-power implementations where the bit depth is re-computed regularly.

\subsection{Energy savings}
 \begin{figure}[t]
	\centering
	\includegraphics[width=8.5cm, height=5cm]{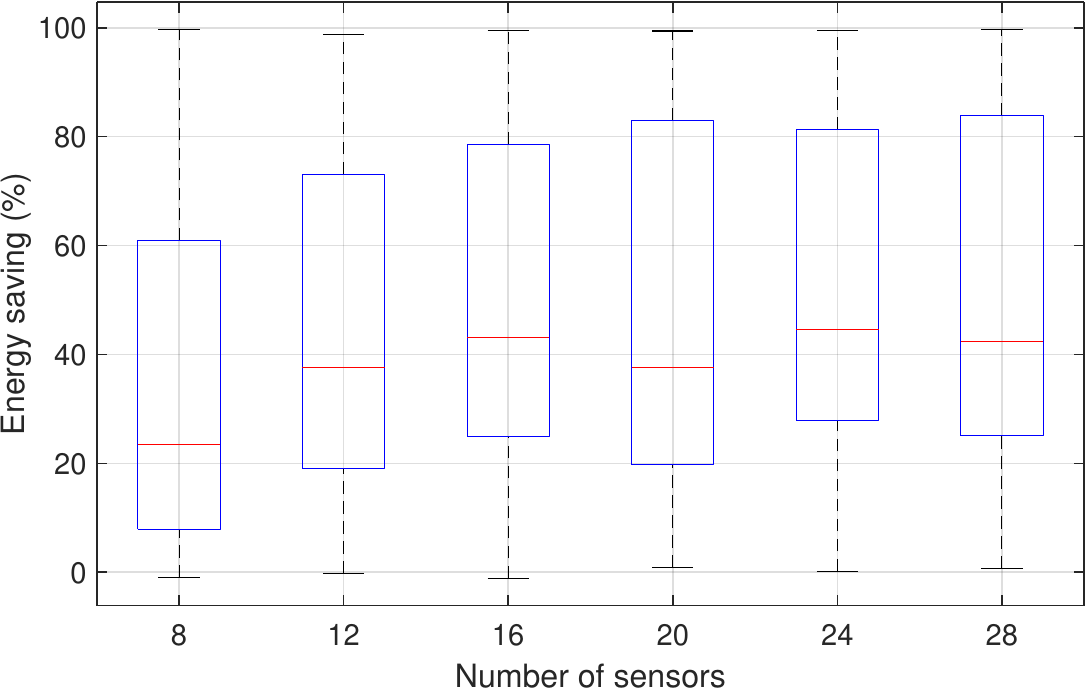}
	% \includegraphics[width=10cm, height=6.5cm]{./img/boxplots_energysaving.pdf}  % Single column
	% For double column [width=8.5cm, height=5cm]
	\caption{Energy savings of the proposed CCP-linMMSE algorithm with respect to
	 homogeneous allocation (same bit depth for every sensor) for several total number of sensors \(K\).}
	\label{fig:energy_saving_ccp}
 \end{figure}
We are now interested in the energy savings with respect to homogeneous allocation, given by
\begin{align}
	\xi_E = 100 \cdot \left ( 1- E_{\text{proposed}} / E_{\text{homog}} \right ) \, ,
\end{align}
where \(E_{\text{proposed}}\) and \(E_{\text{homog}}\) are the total communication energy across all sensors
corresponding respectively to the solution of (\ref{eq:LinearBitAlloc}) given by Algorithm \ref{alg:ccp_bitalloc} and to
homogeneous allocation.
Figure \ref{fig:energy_saving_ccp} shows the energy savings \(\xi_E\) from 100 scenarios with different numbers of
sensors in the WSN and two noise sources. 
In each scenario,  (\ref{eq:LinearBitAlloc}) is solved with the same MMSE threshold \(\eta\) as the corresponding
homogeneous allocation, from 16 to 2 bits.
It can be readily observed that the energy savings are significant in most scenarios, with the lower and upper quartiles
being between 20\% and about 80\% for all different numbers of sensors in the WSN except for 8, where these 
quartiles lie between about 10\% and 60\%.

\subsection{Computation time} \label{subsec:CompTime_CCP}
 \begin{figure}[!t]
    \centering
    \begin{subfigure}[t]{0.6\textwidth}
        \includegraphics[width=8.5cm, height=5cm]{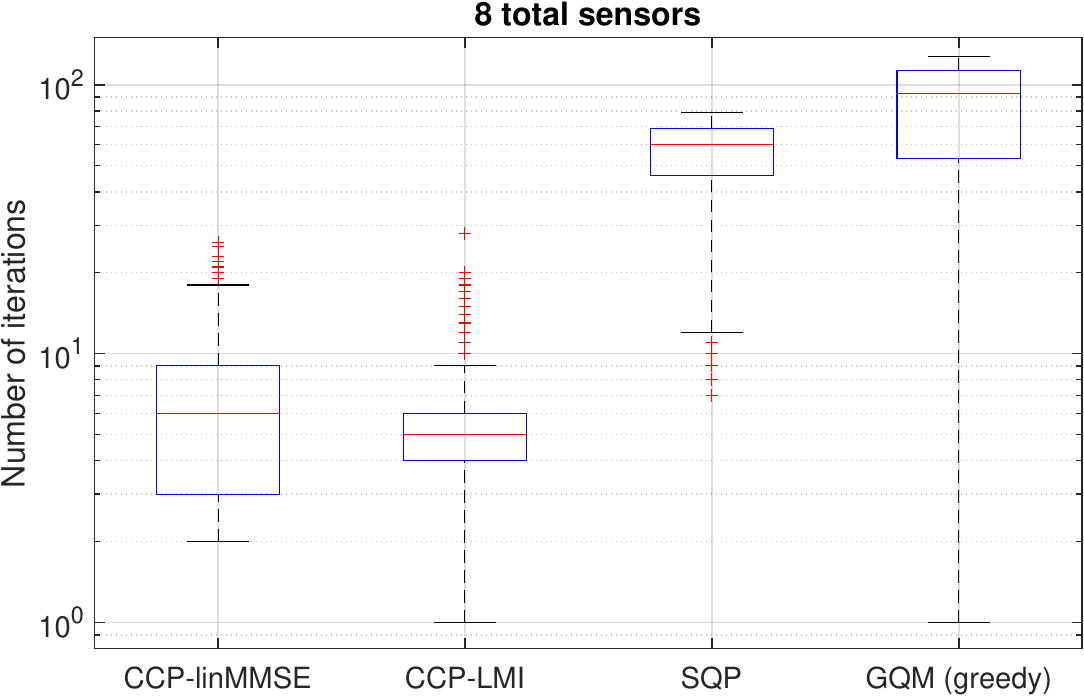}
        \caption{}
        \label{fig:nbiter_8}
    \end{subfigure}
    ~ %add desired spacing between images, e. g. ~, \quad, \qquad, \hfill etc. 
    \begin{subfigure}[t]{0.6\textwidth}
    	\includegraphics[width=8.5cm, height=5cm]{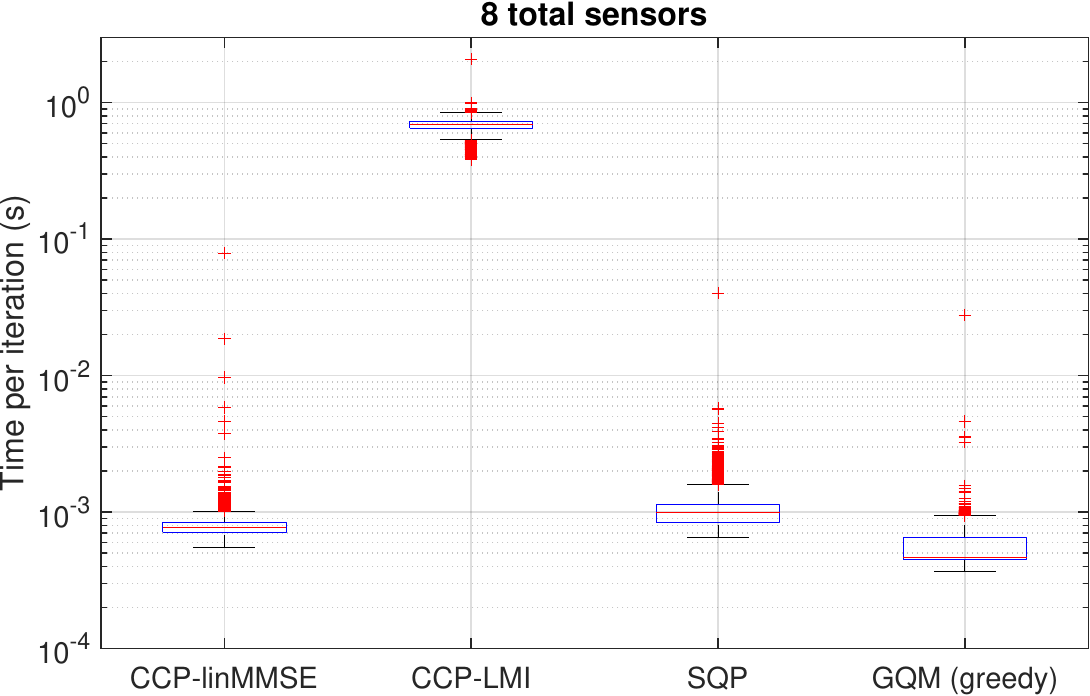}
        \caption{}
        \label{fig:timeiter_8}
    \end{subfigure}
    \caption{Comparison of the number of iterations and average time per iteration needed to reach a solution for 
    CCP-linMMSE, CCP-LMI, SQP and GQM (greedy) for 100 different scenarios with \(K = 8\) sensors in the WSN.}
\end{figure}
 \begin{figure}[!t]
    \centering
    \begin{subfigure}[t]{0.6\textwidth}
    	\includegraphics[width=8.5cm, height=5cm]{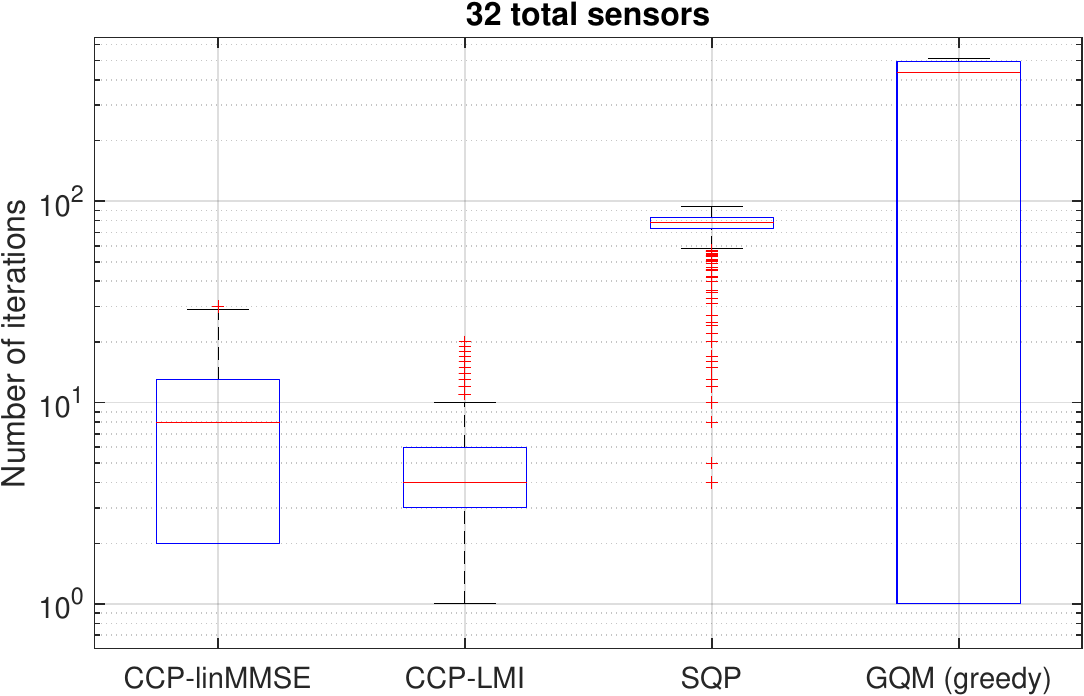}  % Double column
        \caption{}
        \label{fig:nbiter_32}
    \end{subfigure}
    ~ %add desired spacing between images, e. g. ~, \quad, \qquad, \hfill etc. 
    \begin{subfigure}[t]{0.6\textwidth}
    	\includegraphics[width=8.5cm, height=5cm]{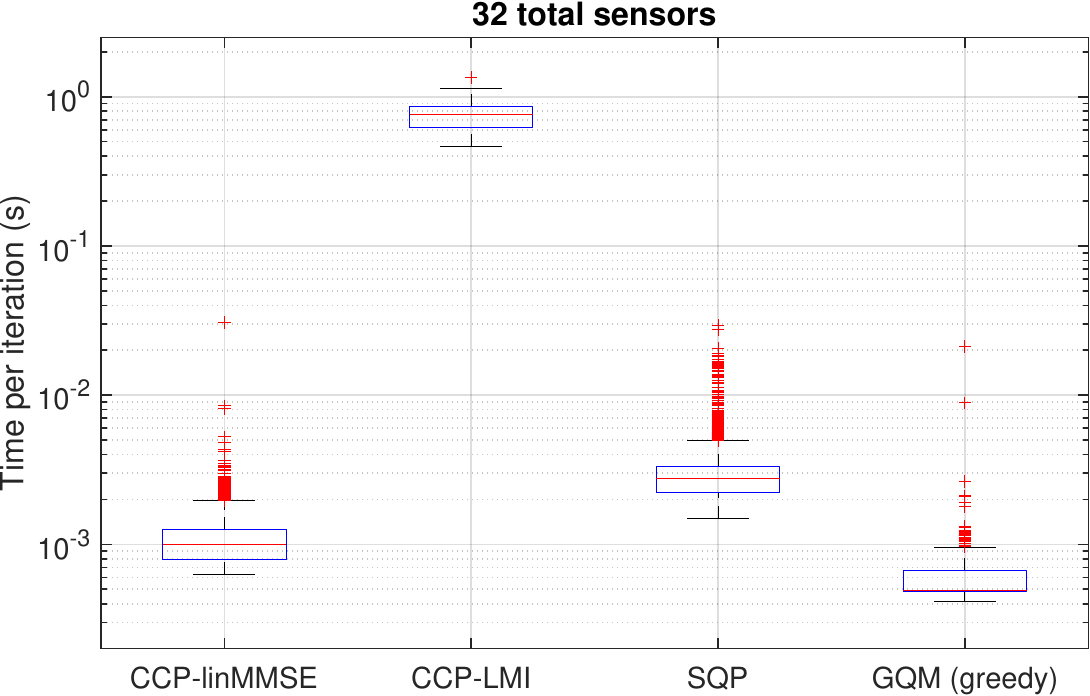}
        \caption{}
        \label{fig:timeiter_32}
    \end{subfigure}
    \caption{Comparison of the number of iterations and average time per iteration needed to reach a solution for 
    CCP-linMMSE, CCP-LMI, SQP and GQM (greedy) for 100 different scenarios with \(K = 32\) sensors in the WSN.}
\end{figure}
Here we compare the time spent in computing a solution of the bit depth allocation problem (\ref{eq:LinearBitAlloc}) 
for the CCP-LMI, SQP and GQM (greedy) methods with our proposed CCP-linMMSE method. 
For this comparison, we measure the number of iterations, the average time per iteration, and the total time spent 
to compute each solution in each of 100 different scenarios, for \(K = 8\) and \(K = 32\) sensors respectively. 
In each scenario the solution is computed for 20 different values of the MMSE threshold \(\eta\).

Figures \ref{fig:nbiter_8} and \ref{fig:nbiter_32} show the number of iterations it takes to reach a solution in each
scenario and each value of the MMSE threshold \(\eta\) for the four methods considered.
It can be readily seen that both methods based on the CCP need in general fewer iterations than the SQP and GQM 
methods.
This observation is true for both \(K = 8\) and \(K = 32\) total sensors in the WSN, and this increase in the number of 
sensors does not significantly affect the number of iterations of the CCP-linMMSE, CCP-LMI and SQP methods, but
it has a clear effect on the greedy method (GQM).
Besides, the GQM has the largest variability in the number of necessary iterations.
Both observations are due to its inherent trade-off between accuracy and speed \cite{DeLaHuchaArce16, DLHA17-01},
where in each step only one bit is changed in order to obtain the best accuracy.
As a result, more iterations are needed compared to the other methods, and this number increases with the number 
of sensors.
A low number of iterations usually happens for lower values of \(\eta\), as this generally means that fewer bits can be
removed before achieving the MMSE threshold.
Regarding the SQP method, it can be observed that the required number of iterations is one order of magnitude 
higher than both methods based on CCP.
The reason of this difference lies in the approach that these methods take. 
SQP solves a sequence of convex problems with a second-order approximation of the cost function and a linear 
approximation of the constraints, and it then needs to perform a line search on a merit function to
limit progress at each iteration. 
By contrast, the convex problem solved in each iteration of a method based on CCP approximates the concave terms,
while the full information of the convex terms is preserved.
Besides, unlike in SQP, limits on progress are not necessary due to the global property of the linear upper bound of the concave terms, in our case the MMSE (\ref{eq:LinearizedMMSE}).

We now turn our attention to  the average time per iteration (as measured in MATLAB) in each scenario and value
of the MMSE threshold \(\eta\) for each of the four methods considered, which is shown in Figures \ref{fig:timeiter_8}
and \ref{fig:timeiter_32}.
It can be clearly observed that the CCP-LMI method's time per iteration is about three orders of magnitude
higher than the time per iteration of the other three methods.
The CCP-LMI method needs to solve a semi-definite program (SDP) in each iteration, which is generally more complex
than the solution of problem (\ref{eq:LinearBitAlloc_Linearized}), using Algorithm \ref{alg:search_dual}, in the 
CCP-linMMSE method, the greedy selection according to the impact metric of the GQM, and the quadratic program of
the SQP method.
With respect to SQP, our proposed CCP-linMMSE method requires a similar amount of time per iteration, although it
scales better with the number of sensors in the network.
This may be due to the fact that CCP-linMMSE does not need a second order approximation of the cost function,
which avoids computing and updating a Hessian matrix, and solving problem (\ref{eq:LinearBitAlloc_Linearized}) is
often simpler than solving a quadratic program.
Finally, the GQM scores the shortest time per iteration of all methods, which is expected since the impact metric for
bit depth changes on each single sensor signal can be efficiently computed from the linear MMSE estimator's 
coefficients \cite{DeLaHuchaArce16, DLHA17-01}.
However, the limitation of steps of 1 bit (for accuracy) on a single sensor signal (for the computational efficiency of
each iteration) result in a large number of iterations in many cases, as we previously explained.

\begin{figure}[t]
	\centering
	\includegraphics[width=8cm, height=5cm]{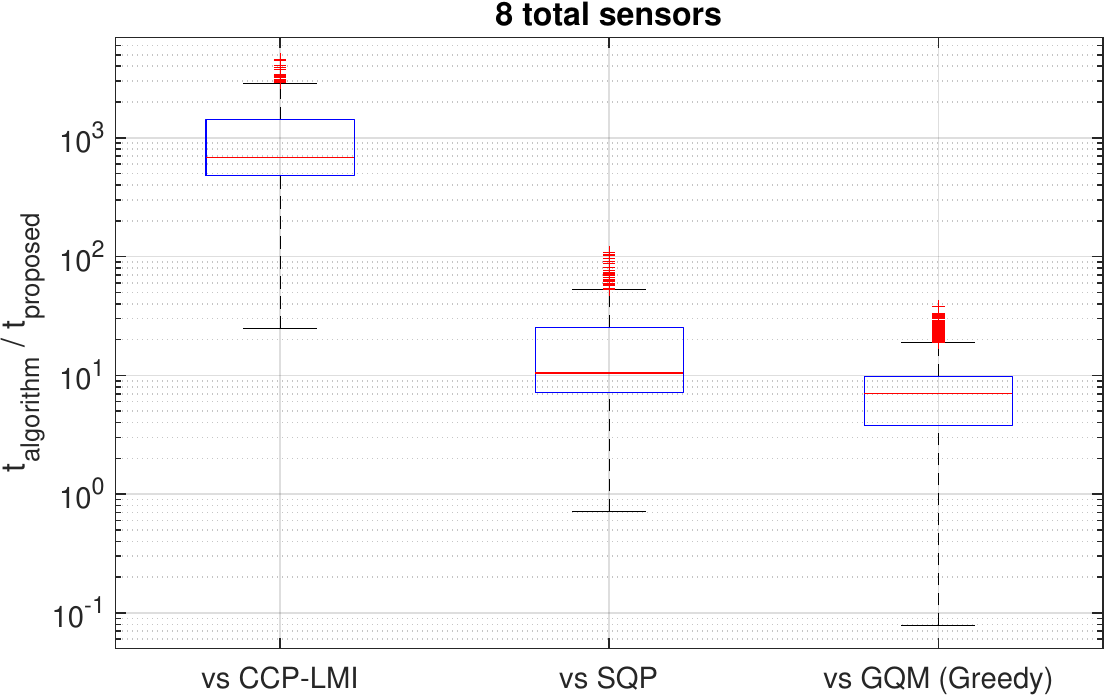}
	% For double column [width=8.5cm, height=5cm]
	% \includegraphics[width=10cm, height=6.5cm]{./img/boxplots_timecomp_8sensors.pdf}
	\caption{Comparison of computation time to reach a solution for CCP-LMI, SQP and GQM (greedy) with respect to 
	our proposed CCP-linMMSE method for 100 different scenarios with \(K = 8\)  sensors in the WSN.}
	\label{fig:comptime_comparison_8sensors}
 \end{figure}
\begin{figure}[t]
	\centering
	\includegraphics[width=8cm, height=5cm]{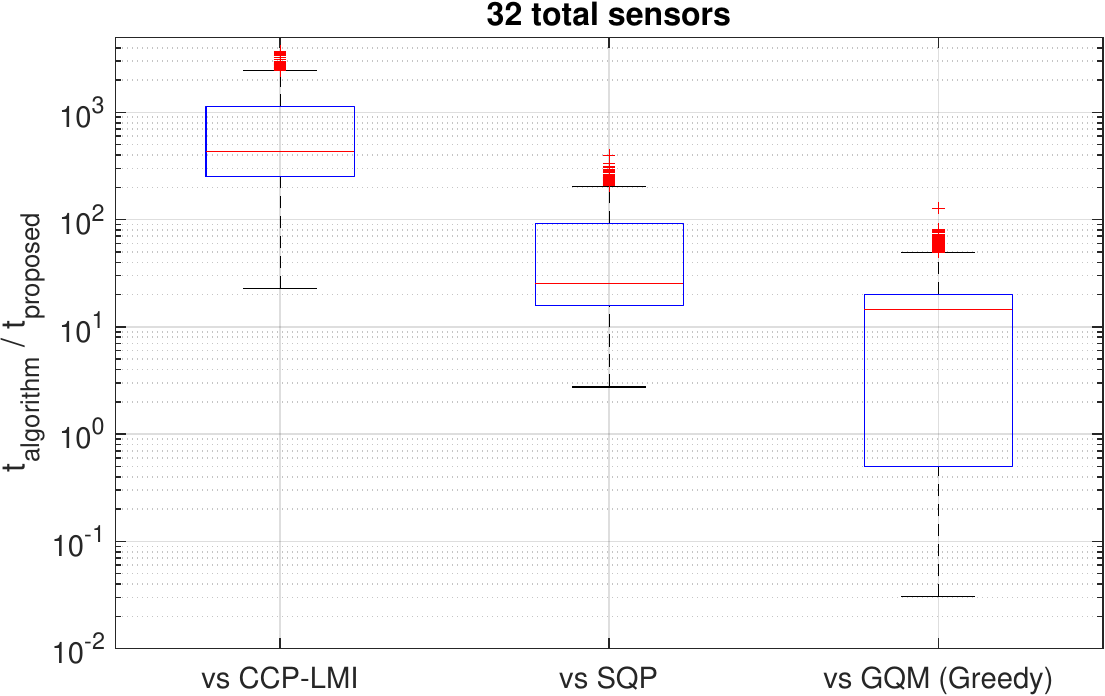}
	% For double column [width=8.5cm, height=5cm]
	% \includegraphics[width=10cm, height=6.5cm]{./img/boxplots_timecomp_32sensors.pdf}
	\caption{Comparison of computation time to reach a solution for CCP-LMI, SQP and GQM (greedy) with respect to 
	our proposed CCP-linMMSE method for 100 different scenarios with \(K = 32\) sensors in the WSN.}
	\label{fig:comptime_comparison_32sensors}
 \end{figure}
In order to compare, for each scenario and value of \(\eta\), the computational efficiency of our proposed 
CCP-linMMSE method against the others, we use as a figure of merit the quotient 
\begin{align} \label{eq:TimeCompFactor}
	t_{\text{factor}} = \frac{t_{\text{algorithm}}}{t_{\text{proposed}}} \, ,
\end{align}
where \(t_{\text{algorithm}}\) is the time spent to reach a solution for any of the three benchmark algorithms, 
and \(t_{\text{proposed}}\) is the time spent to reach a solution for our proposed CCP-linMMSE algorithm.
In Figures \ref{fig:comptime_comparison_8sensors} and \ref{fig:comptime_comparison_32sensors} it can be observed
that our proposed method is in general significantly faster than all three benchmark algorithms.
For the very rare cases where SQP is faster for \(K = 8\) sensors, CCP-linMMSE is almost as fast.
Finally, CCP-linMMSE is also faster than the GQM in at least 80\% of the cases.
Besides, we highlight that CCP-linMMSE has the advantage of being guaranteed to converge to a local minimum, as 
we explained in Section \ref{subsec:ConvergenceBitAlloc}, which is not possible to guarantee for the GQM.

\subsection{Dynamic tracking scenarios}
% Goal and motivation.
We now consider a scenario with a moving source, with the goal of illustrating the performance of the CCP-linMMSE algorithm in a dynamic environment.
Unlike in the static scenarios that we have considered so far, the sensor signals now contain a non-stationary component due to the source's movement.
As a consequence, earlier signal segments become progressively less relevant to the signal estimation task, and therefore to the bit depth allocation problem as well.
In this type of scenario, we will not start the CCP-linMMSE algorithm from scratch for every signal segment, but instead we perform a limited number of iterations starting from the solution to the bit depth allocation of the previous signal segment.
This will allow us to evaluate the tracking capabilities of the algorithm.
\begin{figure}[t]
	\centering
	\includegraphics[width=8cm, height=5cm]{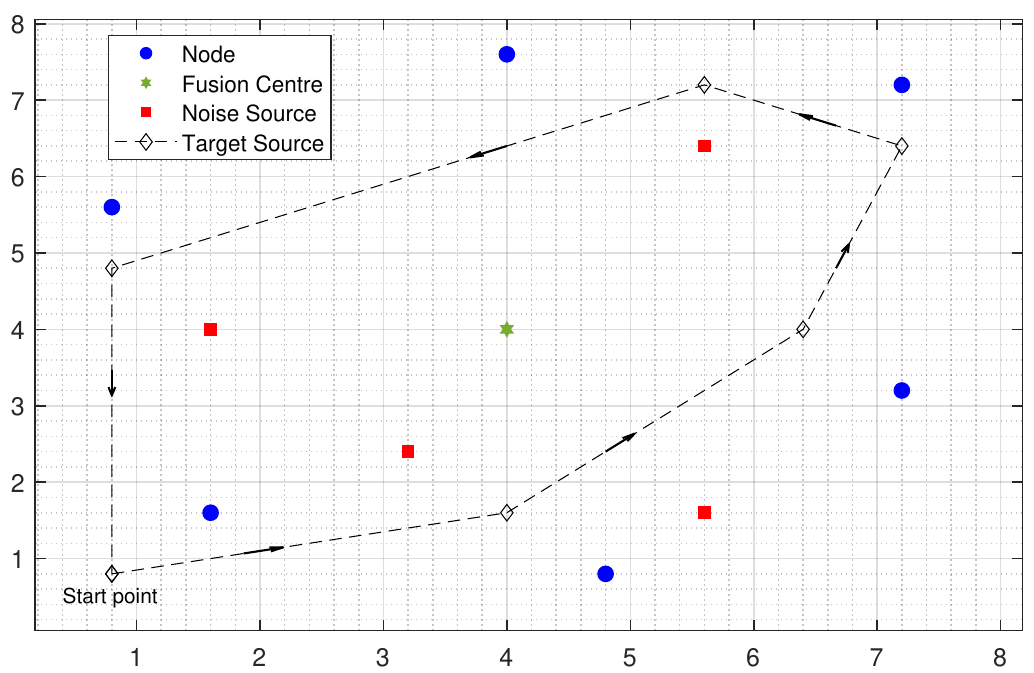}
	% \includegraphics[width=10cm, height=7cm]{./img/Scenario2_Tracking_BitAlloc.pdf}
	% For double column [width=8.5cm, height=5cm]
	\caption{Diagram of the scenario with a moving desired source.}
	\label{fig:scenario_mov_src}
 \end{figure}

% Scenario description
Our scenario consists of a \(8 \times 8 \) m area with six sensor nodes, four noise sources, and a fusion centre, whose positions are depicted in Figure \ref{fig:scenario_mov_src}.
The nature of the (desired and noise) sources' signals, the propagation model and the energy per bit coefficients \(\rho_k\) are the same as described at the beginning of Section \ref{sec:Simulations}.
The desired source is stationary for 1 second at the start point and then describes its trajectory, also shown in Figure \ref{fig:scenario_mov_src}, with a constant speed of 0.5 m/s.
It emits a signal whose average power is six times that of any individual noise source signal, so the input SNR of the sensor signals at the start point lies in the range of (-3, 5) dB.
The sensor nodes acquire their signals with a sampling frequency of \(f_s = 8\) kHz.
These signals are processed in blocks of \(N_{\text{block}}\) samples, and \(\mathbf{Y}(m)\) denotes the \(K \times N_{\text{block}}\) matrix containing the sensor signal samples of the \(m\)-th block.

% Second order statistics estimation via sliding window.
A sliding window of length \(L\) blocks is used to estimate the second order statistics \(\mathbf{R}_{yy}\) and \(\mathbf{r}_{yd}\), i.e.,
\begin{align} \label{eq:Ryy_sliding_window}
	\mathbf{\hat{R}}_{yy}(m) = \mathbf{\hat{R}}_{yy}(m-1) 
	+ \mathbf{\hat{R}}_{yy}^{ \text{(blk)} }(m) 
	- \mathbf{\hat{R}}_{yy}^{ \text{(blk)} }(m - L + 1) \, ,
\end{align}
where \(\mathbf{\hat{R}}_{yy}^{\text{(blk)}}(m) = \frac{1}{N_{\text{block}}} \mathbf{Y}(m) \mathbf{Y}^H(m)\) is the estimated covariance matrix of the \(m\)-th sensor signal block.
An analogous procedure to (\ref{eq:Ryy_sliding_window}) is applied to estimate \(\mathbf{r}_{yd}\)\footnote{As explained at the beginning of Section \ref{sec:Simulations}, knowledge of \(d\) is assumed for the estimation of \(\mathbf{r}_{yd}\), since we are interested in the effects of quantization.}.
For our scenario, we set the values \(N_{\text{block}} = 800\) and \(L =10\), which corresponds to an observation window of 1 second with a 90\% overlap between consecutive windows.
The period where the source remains stationary at the start point is used to initialize \(\mathbf{\hat{R}}_{yy}\) and 
\(\mathbf{\hat{r}}_{yd}\).
During this period, the maximum bit depth \(b_{\text{max}} = 16\) bits is assigned to all sensor signals.

Afterwards, a bit depth allocation problem (\ref{eq:LinearBitAlloc}) is solved using the CCP-linMMSE algorithm every time a new sensor signal block is available.
The solution of this problem when the \(m\)-th block arrives determines the set of bit depths used to quantize the sensor signals in the \((m+1)\)-th block. 
In order to determine the MMSE threshold \(\eta(m)\), i.e., the maximum MMSE allowed for the \(m\)-th bit depth allocation problem, we set a minimum signal-to-error (SER) threshold 
\(\text{SER}_{\text{min}}\).
The SER at the \(m\)-th signal block is given by
\begin{align} \label{eq:SER_def}
	\text{SER}(m) = \frac{\hat{P}_d (m)}{\text{MMSE} (m)}.
\end{align}
 The MMSE threshold \(\eta(m)\) is then determined by \(\text{SER}_{\text{min}}\) and \(\hat{P}_d (m)\), which is an estimate of the desired source's power at \(m\)-th sensor signal block.
 Note that initial points may be infeasible, i.e., \(\text{MMSE}^{(0)}(m) > \eta(m)\), where \(\text{MMSE}^{(j)}(m)\) is the MMSE at the \(j\)-th iteration of the CCP-linMMSE algorithm and the \(m\)-th signal block.
In this case, the solution is set to \(b_k = b_{\text{max}}, \forall k \in \mathcal{K}\). 

To evaluate the behaviour of the CCP-linMMSE algorithm, we will look at both the MMSE predicted by the constraint on the bit depth allocation problem and the actual MMSE between the true and estimated desired signals.
The value of the MMSE constraint function (\ref{eq:MMSE_Function_QPower}) at the solution \(\mathbf{p^*}(m)\) for the \(m\)-th signal block can be expressed as
\begin{align} \label{eq:MMSE_constr_dyn}
	& \text{MMSE}_{\text{constr}} (m) = \nonumber \\
	& \hat{P}_d (m) - \mathbf{\hat{r}}_{yd}^H (m) 
	\left ( \mathbf{\hat{R}}_{y y}(m) + \text{diag}(\mathbf{p^*}(m)) \right )^{-1} \mathbf{\hat{r}}_{yd} (m) \, .
\end{align}
The actual MMSE between the true and estimated desired signals can be computed from its definition as
\begin{align} \label{eq:MMSE_def_dyn}
	\text{MMSE}_{\text{def}} (m) = E \left \{ | d(m) - \hat{d}(m) |^2 \right \} , \, 
\end{align}
where \(d(m)\) and \(\hat{d}(m)\) respectively denote, at the \(m\)-th signal block, the ground truth desired signal and its estimate using the linear MMSE estimator 
\(\mathbf{\hat{w}}_m = \mathbf{\hat{R}}_{y y}^{-1}(m) \mathbf{\hat{r}}_{yd} (m)\).

\begin{figure}[t]
	\centering
	\includegraphics[width=8cm, height=5cm]{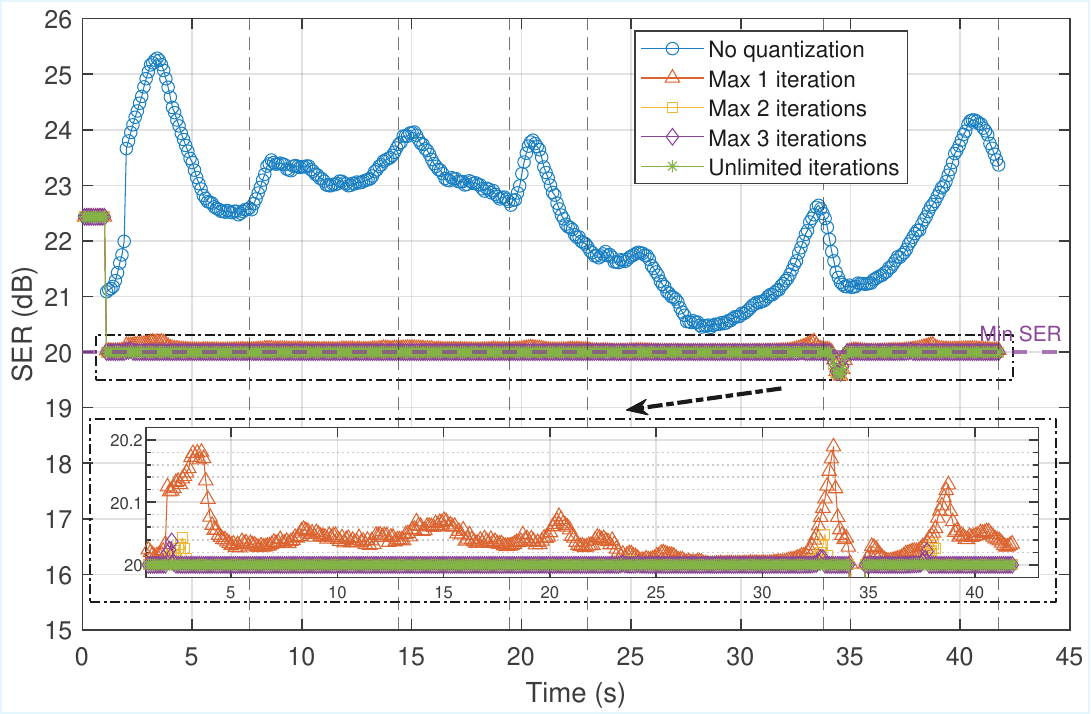}
	% \includegraphics[width=12cm, height=8cm]{./img/SER_sndord_Scen2_minSER_20_with_zoom.pdf}
	% For double column [width=8.5cm, height=5cm]
	\caption{SER (\ref{eq:SER_def}) during the desired source's trajectory and different limits on the number of iterations of the CCP-linMMSE algorithm.
	The MMSE at each signal block is computed through (\ref{eq:MMSE_constr_dyn}).
	The dashed vertical lines mark the arrival at the vertices of the trajectory.}
	\label{fig:ser_sndord_scen2_min20}
 \end{figure}

\begin{figure}[t]
	\centering
	\includegraphics[width=8cm, height=5cm]{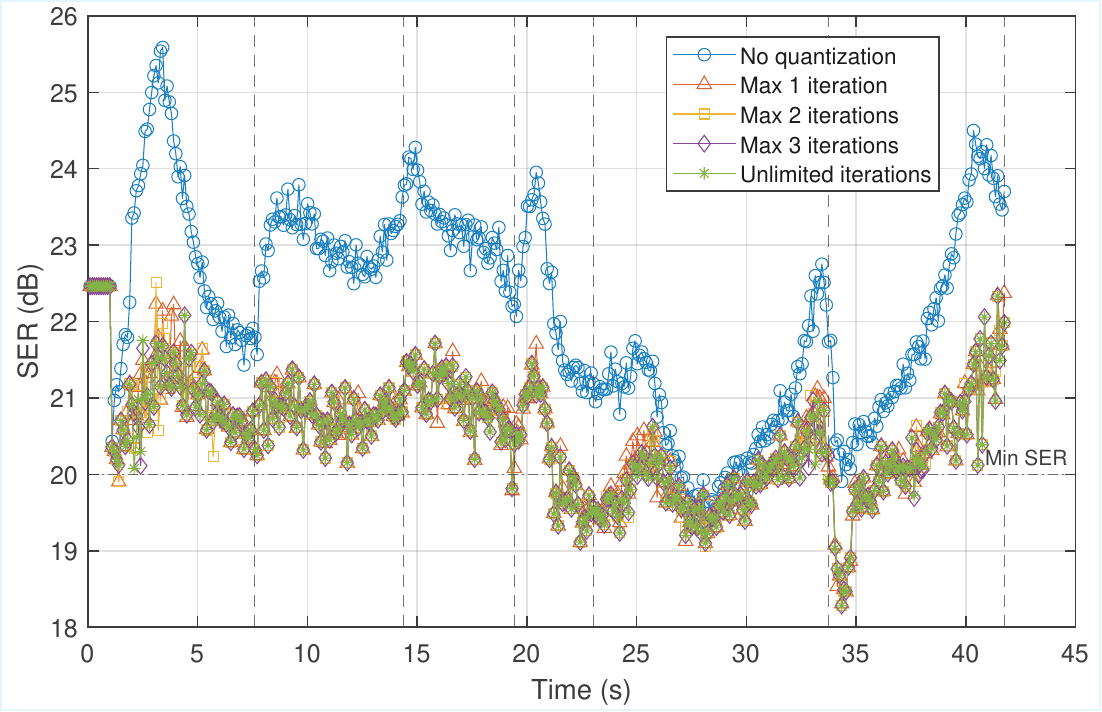}
	% \includegraphics[width=12cm, height=8cm]{./img/SER_def_Scen2_minSER_20.pdf}
	% For double column [width=8.5cm, height=5cm]
	\caption{SER (\ref{eq:SER_def}) during the desired source's trajectory and different limits on the number of iterations of the CCP-linMMSE algorithm.
	The MMSE at each signal block is computed through (\ref{eq:MMSE_def_dyn}).
	The dashed vertical lines mark the arrival at the vertices of the trajectory.}
	\label{fig:ser_def_scen2_min20}
 \end{figure}
 
Figures \ref{fig:ser_sndord_scen2_min20} and \ref{fig:ser_def_scen2_min20} depict the SER at every signal block for the whole duration of the desired source's trajectory, where \(\text{SER}_{\text{min}}\) is set to 20 dB.
The MMSE is computed, respectively, according to (\ref{eq:MMSE_constr_dyn}) and (\ref{eq:MMSE_def_dyn}).
To evaluate the effect of the number of iterations of the CCP-linMMSE algorithm, the bit depth allocation at each signal block is solved with a limit of 1, 2, 3 and unlimited iterations.
As a reference, the SER with no quantization at all is also shown.
It can be observed in Figure \ref{fig:ser_sndord_scen2_min20} that the SERs corresponding to 2, 3 and unlimited iterations stay very close to the \(\text{SER}_{\text{min}}\) target.
The SER corresponding to 1 iteration presents comparatively more deviation from \(\text{SER}_{\text{min}}\), which is expected since in this case the CCP-linMMSE algorithm is not allowed enough time to reach the corresponding target MMSE.
In any case, this deviation is also rather small.
An exception is a short period of time just before \(t = 35\) seconds, where \(\text{SER}_{\text{min}}\) cannot be met, creating a sequence of infeasible initial points which results in a temporary return to the maximum bit depth allocation \(b_{\text{max}} = 16\) bits for all sensor signals.

The SER derived from \(\text{MMSE}_{\text{def}} (m)\) (\ref{eq:MMSE_def_dyn}), shown in Figure \ref{fig:ser_def_scen2_min20}, presents larger deviations around \(\text{SER}_{\text{min}}\).
The reason is that the estimated covariance matrix \(\mathbf{\hat{R}}_{yy}(m)\) already contains not only errors due to the finite window length and the non-stationarity of the signals, but also quantization errors due to the quantized signal blocks \(\mathbf{Y}(m)\).
Besides, the quantization errors do not correspond to the same set of bit depths, since each of the \(L\) blocks used to estimate \(\mathbf{\hat{R}}_{yy}(m)\) is quantized with the solution of their corresponding allocation problem.
Nevertheless, these deviations are not large, and in fact the SER is sometimes slightly better than the value of the MMSE constraint function given by (\ref{eq:MMSE_constr_dyn}).
It is important to highlight the limited difference between the SER achieved by a small number of iterations and that achieved by an unlimited number of iterations, which shows good tracking performance of the time-recursive implementation of the CCP-linMMSE algorithm studied here.
 
\begin{figure}[t]
	\centering
	\includegraphics[width=8cm, height=5cm]{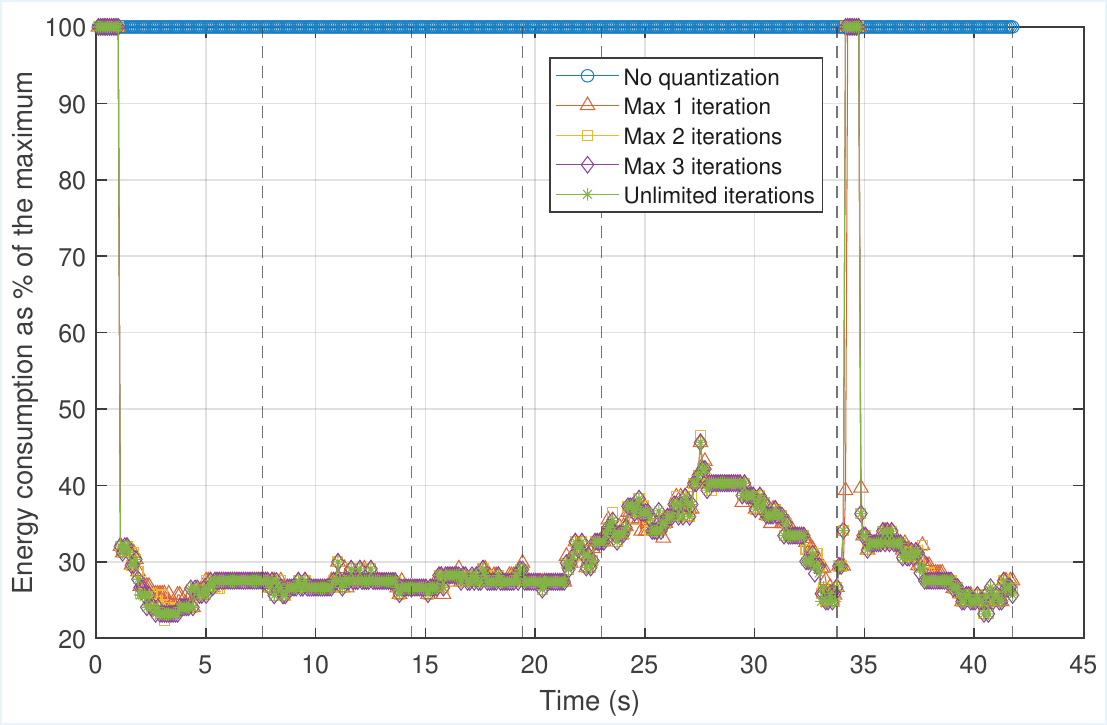}  % Double column
	% \includegraphics[width=12cm, height=8cm]{./img/Energy_perc_Scen2_minSER_20.pdf}  % Single column
	% For double column [width=8.5cm, height=5cm]
	\caption{Energy consumption at every signal block during the desired source's trajectory and different limits on the number of iterations of the CCP-linMMSE algorithm.
	The dashed vertical lines mark the arrival at the vertices of the trajectory.}
	\label{fig:energy_percent_scen2_min20}
 \end{figure}

Finally, Figure \ref{fig:energy_percent_scen2_min20} depicts the energy consumption, as a percentage of the maximum, at every signal block for the whole duration of the desired source's trajectory.
For all maximum iterations allowed of the CCP-linMMSE algorithm, the energy consumption stays between 20\% and 40\% of the maximum, i.e., where all sensor signals are always assigned \(b_{\text{max}} = 16\) bits.
As previously mentioned , there is a short period of time, just before \(t = 35\) seconds, where the algorithm is unable to meet the chosen \(\text{SER}_{\text{min}}\).
It can be observed that the energy consumption jumps to the maximum, due to the temporary return to the maximum bit depth allocation \(b_{\text{max}} = 16\) bits for all sensor signals.

\section{Conclusions} \label{sec:Conclusions}
We have studied a bit depth allocation problem for linear MMSE signal estimation in a WSN where sensors exchange
information through digital communications with orthogonal medium-access control. 
This commonly results in a communication energy model depending linearly on the bit depths used to quantize the
sensor signal samples.
We have shown that, for general scenarios with correlated noise signal components, the (non-convex) bit depth
allocation problem can be formulated as a difference of convex (DC) problem, which allows the use of the 
concave-convex procedure (CCP) to find a local optimum.
For our proposed algorithm, named CCP-linMMSE as it results from the linearization of the MMSE function, we have
shown that the convex problem at each iteration of CCP can be solved with a worst-case complexity equal to the total
number of sensors under mild conditions.
Through simulations, we have also shown that the energy savings resulting from the CCP-linMMSE algorithm are indeed
significant compared to homogeneous allocation.
Additionally, we have assessed that the proposed CCP-linMMSE algorithm is significantly faster than previously
proposed algorithms for bit depth allocation and generic optimization methods, and it scales better with the total
number of sensors.
This computational efficiency results from needing a limited number of inexpensive iterations, while other methods
either need a substantially higher number of iterations to reach a solution than CCP-linMMSE, or are significantly
slower per iteration.
Finally, in a tracking scenario with non-stationary signals, the time-recursive implementation of the CCP-linMMSE algorithm is shown to stay around the optimal bit depth allocation, even when the number of iterations allowed is limited.

% Appendices
\begin{appendices}
\section{Concavity of MMSE as a function of quantization noise power}
\label{app:ConcavityMMSE}
To show that 
\(J(\mathbf{p}) = P_d - \mathbf{r}_{yd}^H \left ( \mathbf{R}_{y y} + \text{diag}(\mathbf{p}) \right )^{-1} \mathbf{r}_{yd} \) 
is concave on its domain \(\text{Dom } J = \{ p_k \geq 0,  k \in \{1, \dots, K\} \} \), we will first show that the function
\(g(\mathbf{p}) = \mathbf{r}_{yd}^H \left ( \mathbf{R}_{y y} + \text{diag}(\mathbf{p}) \right )^{-1} \mathbf{r}_{yd} \) is
convex.
A function \(g\) is convex if and only if its epigraph is a convex set \cite{BoydVandenberghe04}, i.e., 
if \(\text{epi } g = \{(\mathbf{p}, \gamma) : \mathbf{p} \in \text{Dom } g,  g(\mathbf{p}) \leq \gamma \}\) is convex.
The condition \(g(\mathbf{p}) \leq \gamma\) can be written as 
\begin{align} \label{eq:AuxEqConc_1}
	\mathbf{r}_{yd}^H \left ( \mathbf{R}_{y y} + \text{diag}(\mathbf{p}) \right )^{-1} \mathbf{r}_{yd} \leq \gamma \, .
\end{align}
Since all the elements of \(\mathbf{p}\) must be non-negative, \(\mathbf{R}_{y y} + \text{diag}(\mathbf{p})\) is a
positive semi-definite matrix and the Schur complement lemma \cite{BoydVandenberghe04} allows to state the following
condition equivalent to (\ref{eq:AuxEqConc_1}),
\begin{align} \label{eq:AuxEqConc_2}
	\begin{bmatrix}
	 \mathbf{R}_{y y} + \text{diag}(\mathbf{p}) & \mathbf{r}_{yd} \\ 
	 \mathbf{r}_{yd}^H  & \gamma 
	 \end{bmatrix}
	 \succeq 0 \, .
\end{align}
The condition (\ref{eq:AuxEqConc_2}) defines a linear matrix inequality (LMI) in \(\mathbf{p}\) and \(\gamma\), 
which defines a convex set \cite{BoydVandenberghe04}, and thus \(\text{epi } g\) is a convex set.
Since \(g(\mathbf{p})\) is a convex function, it immediately follows that \(J(\mathbf{p}) = P_d - g(\mathbf{p})\) is a 
concave function.

\section{CCP algorithm based on a linear matrix inequality formulation}
\label{app:ccp_lmi}
In \cite{DeLaHuchaArce20} it is shown that the MMSE constraint in (\ref{eq:LinearBitAlloc}),
\begin{align} \label{eq:MMSEConstrApp}
	P_d - \mathbf{r}_{yd}^H \left ( \mathbf{R}_{y y} + \mathbf{R}_{e e} \right )^{-1} \mathbf{r}_{yd} \leq \eta \, ,
\end{align}
can be expressed in the variable \(u_k = 2^{2 b_k}\) as the LMI
\begin{align} \label{eq:MMSE_LMI_2}
	\mathbf{R}_0 + \sum_{k = 1}^K u_k \, \frac{12}{D_k^2} \mathbf{E}_{kk} \succeq 0 \, ,
\end{align}
where the notation \(\mathbf{A} \succeq 0\) indicates that the matrix \(\mathbf{A}\) is positive semi-definite,
\(\mathbf{E}_{kk}\) is the \((K+1) \times (K+1) \) matrix with all zero elements except the \(k\)-th element of
its diagonal, which is equal to one, the \((K+1) \times (K+1) \) matrix \(\mathbf{R}_0\) is given by
\begin{align} \label{eq:R0Def}
	\mathbf{R}_0 = 	
	\begin{bmatrix} \mathbf{R}_{yy}^{-1} & \mathbf{R}_{yy}^{-1} \mathbf{r}_{yd} \\ 
	\mathbf{r}_{yd}^H \mathbf{R}_{yy}^{-1}  & \bar{\eta} \end{bmatrix} \, ,
\end{align}
and \(\bar{\eta} = \eta - \left ( P_d - \mathbf{r}_{yd}^H \mathbf{R}_{yy}^{-1} \mathbf{r}_{yd} \right ) \).
It is well known that LMIs, such as (\ref{eq:MMSE_LMI_2}), define convex sets \cite{BoydVandenberghe04}.
The equivalence between (\ref{eq:MMSEConstrApp}) and (\ref{eq:MMSE_LMI_2}) can be obtained through the 
matrix inversion lemma \cite{HornJohnson13} and the Schur complement lemma \cite{BoydVandenberghe04}, 
and we refer the reader to \cite{DeLaHuchaArce20} for a detailed derivation.
\begin{algorithm}[t]
\caption{CCP-LMI algorithm to solve problem (\ref{eq:LinearBitAlloc}).}
\label{alg:ccp_bitalloc_lmi}
\begin{algorithmic}[1]
	\STATE Initialize \(\mathbf{u}_0\) to be a feasible point of (\ref{eq:BitAllocLinEnergy_u}), \(\varepsilon > 0\), compute
	\(\bar{\eta} = \eta - \left ( P_d - \mathbf{r}_{yd}^H \mathbf{R}_{yy}^{-1} \mathbf{r}_{yd} \right ) \) and form the 
	matrix \(\mathbf{R}_0\) as given in (\ref{eq:R0Def}).
	\STATE \(n := 0\)
	\REPEAT 
		\STATE \textit{Majorization:} 
		Compute \(\bar{E}(\mathbf{u}_n), \nabla \bar{E}(\mathbf{u}_n)\) through (\ref{eq:CostFunctionAppB}) and
		(\ref{eq:GradCostFunAppB}).
		\STATE \textit{Minimization:} 
		Set \(\mathbf{u}_{n+1}\) to the solution of the convex approximate problem (\ref{eq:ccp_subprob_lmi}),
		e.g. through an interior point method \cite{BoydVandenberghe04}.
		\STATE \textit{Update iteration} \(n := n+1\).
 	\UNTIL Convergence criterion is met, e.g. \(\bar{E}(\mathbf{u}_n) - \bar{E}(\mathbf{u}_{n+1}) \leq \varepsilon\).
 	\STATE Convert the solution \(\mathbf{u}^*\) back to integer bit depths, e.g. 
 	\(b_{u,k}^* = \text{round} \left ( 2^{-1} \text{log}_2 u_k^* \right ) \).
\end{algorithmic}
\end{algorithm}

Using (\ref{eq:MMSE_LMI_2}), the problem (\ref{eq:LinearBitAlloc}) can be expressed in the variable 
\(\mathbf{u} = [u_1, \dots, u_K]^T\) as
\begin{align} \label{eq:BitAllocLinEnergy_u}
	\underset{\mathbf{u}}{\text{min }} & \sum_{k=1}^{K} \frac{1}{2} \rho_k \text{log}_2 u_k \nonumber \\
	\text{s.t. } & 
	\mathbf{R}_0 + \sum_{k = 1}^K u_k \, \frac{12}{D_k^2} \mathbf{E}_{kk} \succeq 0 \, , \\
	& 1 \leq u_k \leq 2^{2 b_{\text{max}}}, \; \forall k \in \{1, \dots, K\} \nonumber \, .
\end{align}
Problem (\ref{eq:BitAllocLinEnergy_u}) is non-convex since, while the feasible set is convex, the cost function 
\begin{align} \label{eq:CostFunctionAppB}
	\bar{E}(\mathbf{u}) = \sum_{k=1}^{K} \frac{1}{2} \rho_k \text{log}_2 u_k
\end{align}
is the non-negative sum of concave functions, and hence is concave.
An application of CCP, described in Algorithm \ref{alg:ccp_bitalloc_lmi}, requires to linearize the cost function 
around a point \(\mathbf{u}_n\) at the \(n\)-th iteration, resulting in the convex approximate problem
\begin{align} \label{eq:ccp_subprob_lmi}
	\underset{\mathbf{u}}{\text{min }} & \bar{E}(\mathbf{u}_n) + 
	\nabla \bar{E}(\mathbf{u}_n)^T (\mathbf{u} - \mathbf{u}_n) \nonumber \\
	\text{s.t. } & 
	\mathbf{R}_0 + \sum_{k = 1}^K u_k \, \frac{12}{D_k^2} \mathbf{E}_{kk} \succeq 0 \, , \\
	& 1 \leq u_k \leq 2^{2 b_{\text{max}}}, \; \forall k \in \{1, \dots, K\} \nonumber ,
\end{align}
where 
\begin{align} \label{eq:GradCostFunAppB}
	\nabla \bar{E}(\mathbf{u}_n) = \frac{1}{2 \text{ln}2}[\rho_1 u_{n,1}^{-1}, \dots, \rho_K u_{n,K}^{-1}]^T .
\end{align}
As the cost function is linear, (\ref{eq:ccp_subprob_lmi}) is a semi-definite program (SDP), which can be solved through
interior point methods \cite{BoydVandenberghe04}.
The convergence of Algorithm \ref{alg:ccp_bitalloc_lmi} to a critical point of (\ref{eq:BitAllocLinEnergy_u}) follows from
\cite{Lanckriet09} and a similar discussion to that presented in Section \ref{subsec:ConvergenceBitAlloc}.
Again, once a solution \(\mathbf{u}^*\) to (\ref{eq:BitAllocLinEnergy_u}) is found, it must be transformed back to 
integer bit depths as explained in Section \ref{subsec:BitDepthAlloc}.
The simplest method is rounding, i.e. \(b_{u,k}^* = \text{round}(2^{-1} \text{log}_2 u_k^*)\).

\section{Search algorithm for the optimal dual variable of problem (\ref{eq:LinearBitAlloc_Linearized})}
\label{app:SearchDual}
As explained in Section \ref{subsec:SolApproxCompl}, the optimal dual variable \(\lambda_{n+1}\) can be computed
through (\ref{eq:LambdaExpr}).
This requires the determination of the active constraints at the optimal point, i.e., which sensor indices belong to 
\(\mathcal{K}_{\nu}\) and \(\mathcal{K}_{\mu}\).
While naively checking all possibilities would have a worst-case complexity of \(2^K\) iterations, a similar reasoning to
that of exact algorithms for waterfilling \cite{Palomar03, Palomar05} can be used to achieve a worst-case complexity of 
\(K\) iterations under mild assumptions, as detailed in Algorithm \ref{alg:search_dual}.

Let us assume a specific partition of \(\mathcal{K} = \{1, \dots, K\}\) in the three disjoint sets \(\mathcal{K}_{\nu}\),
\(\mathcal{K}_{\mu}\) and \(\mathcal{K}_I\).
We denote the corresponding value of the dual variable by \(\lambda_{n+1}^{' \text{curr}}\), which can be computed
through (\ref{eq:LambdaExpr}).
Clearly, if \(\lambda_{n+1}^{' \text{curr}}\) satisfies (\ref{eq:Lambda_Cond1}), (\ref{eq:Lambda_Cond2}) and 
(\ref{eq:Lambda_Cond3}) for all the indices that were assumed to belong to \(\mathcal{K}_{\nu}\), \(\mathcal{K}_{\mu}\)
and \(\mathcal{K}_I\), then it is the optimal solution.
If this is not true, then some indices need to be assigned to different sets from those they were assumed to belong.

Let us now assume that an index \(m\) is moved from \(\mathcal{K}_I\) to \(\mathcal{K}_{\nu}\), i.e. 
\(p_m^{\text{new}} = p_{\text{max},m}\).
The new dual variable, denoted by \(\lambda_{n+1}^{' \text{new}}\), is given by
\begin{align} \label{eq:LambdaNew}
	\lambda_{n+1}^{' \text{new}} =  
	\frac{\rho_I - \rho_m}{\eta_I -  \left | w_m(\mathbf{p}_n) \right |^2 p_{\text{max},m}} =
	\gamma \lambda_{n+1}^{' \text{curr}} \, ,
\end{align}
where 
\begin{align} \label{eq:GammaLambda}
	\gamma = 
	\frac{1 - \frac{\rho_m}{\rho_I}}
	{1 - \frac{\left | w_m(\mathbf{p}_n) \right |^2 p_{\text{max},m}}{\eta_I}} \, .
\end{align}
Note that \(\gamma \leq 1\), and hence \(\lambda_{n+1}^{' \text{new}} \leq \lambda_{n+1}^{' \text{curr}}\), if and only if
\begin{align} \label{eq:GammaLeq1}
	\frac{\rho_m}{\rho_I} \geq  \frac{\left | w_m(\mathbf{p}_n) \right |^2 p_{\text{max},m}}{\eta_I} \, .
\end{align}
Rearranging the terms of (\ref{eq:GammaLeq1}) and using (\ref{eq:LambdaExpr}) and 
(\ref{eq:Lambda_Cond2}) we obtain
\begin{align} \label{eq:LambdaCurr_1}
	\lambda_{n+1}^{' \text{curr}} \leq \lambda_{\nu m} \, .
\end{align}
We can thus establish that 
\begin{subequations}
\begin{align} 
	\label{eq:LambdaIFF_nu1} 
	& \lambda_{n+1}^{' \text{curr}} \leq \lambda_{\nu m} \quad \text{if and only if}  \quad 
	\lambda_{n+1}^{' \text{new}} \leq \lambda_{n+1}^{' \text{curr}} \, , \\
	\label{eq:LambdaIFF_nu2} 
	 & \lambda_{n+1}^{' \text{curr}} > \lambda_{\nu m} \quad \text{if and only if}  \quad 
	\lambda_{n+1}^{' \text{new}} > \lambda_{n+1}^{' \text{curr}} \, .
\end{align}
\end{subequations}
An index \(m\) assumed to belong to \(\mathcal{K}_I\) and for which 
\(\lambda_{n+1}^{' \text{curr}} \leq \lambda_{\nu m}\) violates (\ref{eq:Lambda_Cond3}).
From (\ref{eq:LambdaIFF_nu1}) we conclude that said index \(m\) must be moved to \(\mathcal{K}_{\nu}\), as this implies that (\ref{eq:Lambda_Cond2}) is satisfied since 
\(\lambda_{n+1}^{' \text{new}} < \lambda_{n+1}^{' \text{curr}}\), and thus 
\(\lambda_{n+1}^{' \text{new}} \leq \lambda_{\nu m}\), and now \(m\) belongs to \(\mathcal{K}_{\nu}\) instead of
\(\mathcal{K}_I\).
Moving from \(\mathcal{K}_I\) to \(\mathcal{K}_{\nu}\) an index \(m\) for which 
\(\lambda_{n+1}^{' \text{curr}} > \lambda_{\nu m}\)  would result in a contradiction due to (\ref{eq:LambdaIFF_nu2}), 
as then \(\lambda_{n+1}^{' \text{new}} > \lambda_{n+1}^{' \text{curr}}\), which would now violate
(\ref{eq:Lambda_Cond2}).

An analogous reasoning can be followed for an index \(m\) moved from \(\mathcal{K}_I\) to \(\mathcal{K}_{\mu}\) 
to obtain
\begin{subequations}
\begin{align} 
	\label{eq:LambdaIFF_mu1} 
	& \lambda_{n+1}^{' \text{curr}} < \lambda_{\mu m} \quad \text{if and only if} \quad 
	\lambda_{n+1}^{' \text{new}} < \lambda_{n+1}^{' \text{curr}} \, , \\
	\label{eq:LambdaIFF_mu2} 
	 & \lambda_{n+1}^{' \text{curr}} \geq \lambda_{\mu m} \quad \text{if and only if} \quad 
	\lambda_{n+1}^{' \text{new}} \geq \lambda_{n+1}^{' \text{curr}} \, .
\end{align}
\end{subequations}
Note that, as \(\lambda_{\nu m} < \lambda_{\mu m} \, \forall m \in \mathcal{K}\), indices which have been moved to
\(\mathcal{K}_{\nu}\) cannot possibly belong to \(\mathcal{K}_{\mu}\), and viceversa.

In what follows, we assume that
\begin{align} \label{eq:LambdaNuMuCond}
	\lambda_{\nu j} < \lambda_{\mu m} \, , \forall j, m \in \mathcal{K} ,
\end{align}
is satisfied. 
Later on (see (\ref{eq:LambdaNuMu_2})), we motivate that this assumption is typically satisfied in practice.
Under this condition, it follows that either 
\(\lambda_{n+1}^{' (K)} < \underset{k}{\text{min}} \, \lambda_{\mu k}\) or 
\(\lambda_{n+1}^{' (K)} > \underset{k}{\text{max}} \, \lambda_{\nu k}\), where
\begin{align} \label{eq:LambdaInactConstr}
	\lambda_{n+1}^{' (K)} = \frac{\sum_{k = 1}^K \rho_k}{\bar{\eta}_n}
\end{align}
is the dual variable (\ref{eq:LambdaExpr}) assuming \(\mathcal{K}_I = \mathcal{K}\).
This can be explained as follows.
In the first case, none of the constraints in (\ref{eq:Lambda_Cond1}), i.e. those associated with \(K_{\mu}\), are active.
Furthermore, (\ref{eq:LambdaIFF_nu1}) guarantees that \(\lambda_{n+1}^{' \text{new}}\) will only decrease when 
any index \(m\) assumed to belong to \(\mathcal{K}_I\) but which satisfies 
\(\lambda_{n+1}^{' \text{curr}} \leq \lambda_{\nu m}\)
is moved to \(\mathcal{K}_{\nu}\), which happens at each iteration of the search algorithm.
Therefore, we know that the constraints (\ref{eq:Lambda_Cond1}) are not active at the optimal solution (and indeed
at any iteration of the search).
The search finds the optimal \(\lambda_{n+1}'\) in at most \(K\) iterations, when either (\ref{eq:Lambda_Cond2}) and
(\ref{eq:Lambda_Cond3}) are satisfied simultaneously or \(\mathcal{K}_{\nu} = \mathcal{K}\).
An analogous reasoning can be followed for the second case, where none of the constraints in 
(\ref{eq:Lambda_Cond2}), i.e. those associated with \(K_{\nu}\), are active.
Here (\ref{eq:LambdaIFF_mu2}) guarantees that \(\lambda_{n+1}^{' \text{new}}\) would only increase when any index
\(m\) assumed to belong to \(\mathcal{K}_I\) but which satisfies \(\lambda_{n+1}^{' \text{curr}} \geq \lambda_{\mu m}\)
is moved to \(\mathcal{K}_{\mu}\).
The search algorithm resulting from this reasoning is detailed in Algorithm \ref{alg:search_dual}.

Using (\ref{eq:Lambda_Cond1}), (\ref{eq:Lambda_Cond2}) and the definitions of \(p_{\text{max},j}\) and 
\(p_{\text{min},k}\) from (\ref{eq:LinearBitAlloc_2}), the condition (\ref{eq:LambdaNuMuCond}) can also be written as
\begin{align} \label{eq:LambdaNuMu_2}
	\frac{\rho_j}{\left | w_j(\mathbf{p}_n) \right |^2} < 
	\frac{\rho_m}{\left | w_m(\mathbf{p}_n) \right |^2} \left ( \frac{D_j}{D_m} \right )^2 2^{2 b_{\text{max}}} \, , 
	\forall j, m \in \mathcal{K}.
\end{align}
As can be readily seen from (\ref{eq:LambdaNuMu_2}), the satisfaction of (\ref{eq:LambdaNuMuCond}) is indeed
scenario-dependent.
Nevertheless, in most practical situations \(b_{\text{max}}\) is large enough (e.g. \(b_{\text{max}} \geq 8\)) to guarantee
its satisfaction.

If (\ref{eq:LambdaNuMuCond}) is not satisfied, the worst-case complexity of the search depends on how many indices
violate this condition. 
The search algorithm then needs to perform an exhaustive search over the indices for which 
\(\lambda_{\nu j} \geq \lambda_{\mu m}\), until either \(\lambda_{n+1}^{' \text{curr}}\) satisfies (\ref{eq:Lambda_Cond1}), 
(\ref{eq:Lambda_Cond2}) and (\ref{eq:Lambda_Cond3}), in which case it is the optimal dual variable, or exclusively
constraints of either the form (\ref{eq:Lambda_Cond1}) or (\ref{eq:Lambda_Cond2}) remain.
In the latter case, Algorithm \ref{alg:search_dual} can be used again.
Note that, even if (\ref{eq:LambdaNuMuCond}) is not satisfied, if the problem satisfies 
\(\lambda_{n+1}^{' (K)} < \underset{k}{\text{min}} \, \lambda_{\mu k}\) or 
\(\lambda_{n+1}^{' (K)} > \underset{k}{\text{max}} \, \lambda_{\nu k}\), then Algorithm \ref{alg:search_dual} will also
find the optimal dual variable with a worst case complexity of \(K\) iterations.

\section{Satisfaction of a constraint qualification for the bit depth allocation problem (\ref{eq:LinearBitAlloc_2})}
\label{app:LICQ}
The linear independence constraint qualification (LICQ) holds at a specific point if the gradients of the active
constraint functions at that point are linearly independent.
For a point \(\mathbf{p}^*\) satisfying the KKT conditions (\ref{eq:PFTr1})-(\ref{eq:LS_Tr}) we denote the subsets
of the set of active constraints \(\mathcal{K}_{\nu}^* = \{k \in \mathcal{K} : p_k^* = p_{\text{max},k}\}\) and 
\(\mathcal{K}_{\mu}^* = \{k \in \mathcal{K} : p_k^* = p_{\text{min},k}\}\).
The constraints corresponding to these subsets cannot be satisfied simultaneously, so \(\mathcal{K}_{\nu}^* \) and
\(\mathcal{K}_{\mu}^*\) are disjoint sets.
The gradient of the constraint functions \(p_{\text{min},k} -  p_k\)  and \(p_k - p_{\text{max},k}\) are \(-\mathbf{e}_k\)
and \(\mathbf{e}_k\) respectively, where \(\mathbf{e}_k\) is the \(K \times 1 \) vector whose \(k\)-th entry is one and all
other entries  are zero.
We note that these gradients are always linearly independent regardless of how many of the constraints are active. 

The only way for the LICQ to not be satisfied is when the first constraint is active, i.e., \(J(\mathbf{p}^*) = \eta\),
and where the indices in \(\mathcal{K}_{\nu}^* \) and \(\mathcal{K}_{\mu}^*\) correspond exactly to the non-zero
elements of \(\nabla J(\mathbf{p}^*)\).
In that case, \(\nabla J(\mathbf{p}^*)\) can be expressed as a linear combination of the vectors \(\mathbf{e}_k\) and
\(-\mathbf{e}_k\), \(\forall k \in \mathcal{K}_{\mu}^*\) and \(k \in \mathcal{K}_{\nu}^*\).
Since we assumed \(\mathbf{p}^*\) satisfies the KKT conditions, for any index \(k \in \mathcal{K}\) such that 
\(\nabla J(\mathbf{p}^*)_k = 0\) equation (\ref{eq:LS_Tr}) becomes
\begin{align} \label{eq:ZeroElemLS}
	\nabla \tilde{E}(\mathbf{p}^*)_k - \mu_k^* + \nu_k^* = 0 \, .
\end{align}
As \(\nabla \tilde{E}(\mathbf{p}^*)_k = \frac{- \rho_k}{2 \text{ln} 2} (p_k^*)^{-1}\) never vanishes and is always negative, 
the only way to satisfy (\ref{eq:ZeroElemLS}) is when \(\nu_k^* > 0\).
Together with (\ref{eq:CSTr2}), (\ref{eq:CSTr3}) and \(p_{\text{max},k} \neq p_{\text{min},k}\), this implies that 
\(\mu_k^* = 0\), and thus \(k \in \mathcal{K}_{\nu}^*\).

As a conclusion, the LICQ is not satisfied for points \(\mathbf{p}^*\) satisfying the KKT conditions of 
(\ref{eq:LinearBitAlloc_2}) if the two following conditions are satisfied simultaneously:
\begin{enumerate}
	\item \(J(\mathbf{p}^*) = \eta\) 
	\item 
	\begin{enumerate} 
		\item If all elements of \(\nabla J(\mathbf{p}^*)\) are non-zero: 
			\(\mathcal{K}_{\mu}^* \cup  \mathcal{K}_{\nu}^* = \mathcal{K}\), i.e., either \(p_k^* = p_{\text{min},k}\) or
			\(p_k^* = p_{\text{max},k}, \forall k \in \mathcal{K}\).
		\item If some elements of \(\nabla J(\mathbf{p}^*)\) are zero: 
			\(\mathcal{K}_{\mu}^* \cup  \mathcal{K}_{\nu}^* = \mathcal{K}\) and \(\mathcal{K}_{\nu}^*\) contains all the indices of the zero elements of \(\nabla J(\mathbf{p}^*)\).
	\end{enumerate}
\end{enumerate}
Note that it is rather uncommon for the LICQ not to hold, and that it is easy to check whether it is satisfied.
Besides, the LICQ is a sufficient but not necessary condition that qualifies a point satisfying the KKT conditions
for local optimality.
If the LICQ does not hold at a specific point, other constraint qualifications might be satisfied, and we refer the reader
to \cite{Nocedal06} for further details.
If no constraint qualification is satisfied, the point may not be a stationary point.
\end{appendices}

% Bibliography
\bibliographystyle{IEEEtran}
\bibliography{./bibtex/bibliography_bitallocation}

\end{document}